\documentclass[%
reprint,prb,
superscriptaddress,
amsmath,amssymb,
aps,floatfix
]{revtex4-2}
\usepackage{graphicx} 
\usepackage{natbib}

\usepackage{braket}	
\usepackage{xcolor}
\usepackage{appendix}

\newcommand{\vex}[1]{\mathbf{#1}}

\newcommand{\kb}{\mathbf{k}}

\newcommand{\ww}{\omega}
\DeclareMathOperator{\sgn}{sgn}
\DeclareMathOperator{\re}{Re}
\newcommand{\msf}[1]{\mathsf{#1}}

\begin{document}
		\title{Quantum Fisher information in a quenched $p + ip$ superfluid}	
        \author{Stelio Varrone}
        \author{Yunxiang Liao}
        \email{liao2@kth.se}
    \affiliation{Department of Physics, KTH Royal Institute of Technology, SE-106 91
    Stockholm, Sweden}

		\begin{abstract}
The quantum Fisher information (QFI) is widely used to characterize quantum phases and transitions, but its diagnostic power sometimes relies on selecting special generators based on prior knowledge of the underlying physics. We ask whether this requirement can be relaxed in nonequilibrium systems by studying the QFI of the long-time asymptotic state of a two-dimensional $p+ip$ superfluid following an instantaneous quench of the coupling strength, using the particle number within a large subextensive subsystem as the generator. In equilibrium, the ground-state QFI is continuous across the topological transition between the weak-pairing BCS and the strong-pairing BEC phases, showing no direct signature of the transition. After the quench, however, the QFI associated with the same generator distinguishes the three dynamical phases and  can encode the topology of the pre-quench state. In phase I, with a vanishing order parameter, the QFI Fourier spectrum consists of a single  zero-frequency spike determined by the nonequilibrium distribution function. In phase II, with a constant nonzero order parameter, the QFI spectrum contains a zero-frequency spike and two continua separated by a gap set by the minimum asymptotic quasiparticle energy. The  continuum edge behavior reveals whether this minimum occurs at zero or finite momentum. In the former case, the spectral weight vanishes at the edge, with the sign just inside the continuum encoding the pre-quench topology for large subsystem. In phase III, with a time-periodic order parameter, the QFI spectrum exhibits discrete peaks at integer multiples of the oscillation frequency,  together with continua associated with the Floquet quasienergy spectrum. Our results show that driving the system out of equilibrium can enhance the diagnostic power of the QFI for a standard physical observable, revealing information inaccessible in equilibrium.
		\end{abstract}
		
		\maketitle

        \section{Introduction}

			The quantum Fisher information (QFI)~\cite{Liu_2019,Pezze-RMP,PETZ_2011} has recently emerged as a powerful probe of many-body quantum systems, offering valuable insights into their equilibrium properties and nonequilibrium dynamics from a quantum information perspective. 
			Originally introduced in quantum metrology, the QFI bounds the precision with which a parameter can be estimated in quantum measurements through the quantum Cram\'er-Rao bound~\cite{helstrom1969,Caves1994,cramer1999,Metrology}. 
			Beyond its role in quantum metrology, QFI has also been widely used as a witness of multipartite entanglement~\cite{Pezz__2009,Hyllus_2012,toth2012,Si-2}. 
			Violation of an appropriate QFI bound with suitably chosen generators provides a lower bound on the number of particles that must be entangled with one another in a many body quantum system.
		 	Moreover, for thermal states under unitary parametrization, the QFI can be expressed as a weighted integral of the dynamical susceptibility~\cite{Hauke_2016}, which can be measured experimentally using neutron scattering~\cite{neutron} and Bragg spectroscopy~\cite{Bragg-1,Bragg-2}.
			This relation makes the QFI an experimentally accessible tool for studying entanglement~\cite{neutron-1,neutron-2,Si-2,Si-3}. 
			
		Because of its close connection to  entanglement and correlations,   the QFI can, with suitably chosen generators, characterize quantum phases and diagnose quantum phase transitions~\cite{QFI-criticality-0,QFI-criticality,QFI-criticality-2,QFI-topo-1,QFI-topo-2,Liu_2013,Si-1,Si-2,Si-3,QFI-BCS}.
		However, the physical information revealed by the QFI appears to depend crucially on the choice of  generator~\cite{zhang2018,QFI-topo}.
		In many studies~\cite{QFI-topo,Hauke_2016,Liu_2013,zhang2018,Dellanna2023,QFI-criticality-2}, the generators are often tailored to the underlying physics of the system, such as the relevant order parameter, symmetry or topological properties. 
		Such choices  usually require some prior knowledge of the system being probed, limiting the power of the QFI as a diagnostic tool. 
		The limitation becomes particularly relevant for topological phase transitions, which are not generally characterized by a conventional local order parameter~\cite{Topo-Rev}, and their diagnosis using the QFI sometimes relies on specially designed generators that are not simple measurable local operators~\cite{QFI-topo,Dellanna2023}.
		This raises the question of whether the QFI can reveal useful information about the system under study without relying on specially designed generators?

		A natural possibility is to change the state rather than the generator, by driving the system out of equilibrium. In particular, quantum quenches have been shown to reveal information that remains hidden in equilibrium~\cite{PRB,PRA,quench-topo}, suggesting that they may also provide a possible approach to overcoming this limitation of the QFI.	
		The QFI has been investigated in nonequilibrium systems~\cite{QFI-DPT,QFI-DPT2,QFI-ergodic,pappalardi2018,Pappalardi2020}, in particular following quantum quenches~\cite{Srednicki,pappalardi2017}. Previous studies of nonequilibrium QFI have often considered the QFI associated with generators that are already known to reveal the relevant physics (for example, as witnesses of multipartite entanglement or probes of quantum phases) in equilibrium and investigated how it evolves after a quantum quench. Here we ask a different question. We consider the QFI with respect to a generic physical observable that may not reveal the relevant physics in equilibrium, and ask whether it becomes informative following a quantum quench.
			
		The quenched two-dimensional (2D) $p+ip$ superfluid~\cite{PRB,PRA,PRL,QuenchRev,Zhao2009} offers an ideal platform for addressing this question.
		In equilibrium, it exhibits a topological quantum phase transition between the weak pairing topological BCS phase and the strong pairing trivial BEC phase~\cite{Read,Alicea_2012,Nayak_2008}.
		Following an instantaneous quench of the coupling strength, the system exhibits a rich nonequilibrium phase diagram~\cite{PRB,PRL,Levitov,Yuzbashyan2006,Chou_2017}: depending on the pairing strengths before and after the quench, the order parameter either decays to zero in phase I, approaches a constant in phase II, or undergoes periodic oscillations in phase III. 
		These dynamical phases also possess distinct topological properties. Phase I realizes a gapless state, phase II contains both topological and trivial regimes distinguished by the presence or absence of Majorana edge modes in the asymptotic quasiparticle spectrum~\cite{PRB}, and phase III realizes a self-generated Floquet  topological superfluid supporting Majorana edge states~\cite{PRL}.
		Moreover, the mean field dynamics of this model are integrable~\cite{BCS-MF,PRB,yuzbashyan2005,Yuzbashyan06,yuzbashyan2005,Levitov}, allowing the long-time asymptotic QFI to be determined analytically. 
		The model is also experimentally relevant to ultracold atomic gases near $p$-wave Feshbach resonances~\cite{Gurarie2005,Gurarie2007,Feshbach-1,Feshbach-2}, and its dynamical phases have been proposed to be simulated using trapped ion quantum simulators~\cite{PRXQuantum}. 

        In this paper, we investigate the QFI of the long-time asymptotic state of a 2D  $p+ip$ superfluid following an instantaneous quench of the coupling strength, using the particle number within a large and subextensive subsystem as the generator. 
        In equilibrium, the ground state QFI associated with this subsystem particle number operator is continuous across the topological phase transition and does not provide a direct signature of the equilibrium topology. 
        By contrast, an instantaneous quench of coupling strength enables the QFI with the same generator to reveal information that is inaccessible in equilibrium.
        In particular, in dynamical phase I, the QFI is time independent and is determined solely by the nonequilibrium Cooper-pair distribution function.
        In phase II, the QFI becomes time dependent, and its Fourier spectrum consists of a discrete peak at zero frequency, together with two continua separated by a gap, whose width is determined by the minimum of the asymptotic quasiparticle energy dispersion. 
        At the edges of the continua, the QFI Fourier spectrum either vanishes or exhibits a van Hove singularity, depending on whether the minimum of the asymptotic quasiparticle dispersion occurs at zero or finite momentum. 
        More importantly, for phase II quenches and sufficiently large subsystems, when the QFI Fourier spectrum vanishes at the continuum edge, the sign of the QFI spectrum immediately inside the continuum close to the edge, reveals the topology of the pre-quench state, even though the QFI is evaluated in the post-quench asymptotic state. 
        Finally, in the Floquet phase III, the QFI reflects the underlying Floquet dynamics through a series of evenly spaced discrete peaks at integer multiples of the order-parameter oscillation frequency, as well as continua associated with the Floquet quasienergy spectrum.  
        These characteristic feactures in the QFI Fourier spectrum distinguish the three dynamical phases, and reveal the topology of the pre-quench state for a wide class of phase II quenches.
            
        The remainder of the paper is organized as follows. In Sec.~\ref{sec:background}, we briefly review the concept of the QFI, and the 2D $p+ip$ superfluid, focusing on the equilibrium and nonequilibrium properties of the $p+ip$ superfluid relevant to the present work.
        In Sec.~\ref{sec:results},  we present our results for the QFI of the 2D $p+ip$ superfluid in equilibrium and  in the three nonequilibrium dynamical phases following a quantum quench. 
        We also discuss the large subsystem approximation used in the QFI calculation, its regime of validity, and the behaviors of QFI beyond this approximation.
        Finally, in Sec.~\ref{sec:conclusion}, we conclude by summarizing our main results and discussing open questions and future directions.
            
		\section{Background}~\label{sec:background}

        In this section, we review the theoretical background used throughout this work. We begin with a brief overview of the QFI. For a more comprehensive discussion, see Refs.~\cite{Liu_2019,Pezze-RMP} and references therein. We then review the 2D $p+ip$ superfluid, summarizing its equilibrium properties and the nonequilibrium quench dynamics following an interaction quench.
        More details can be found in Refs.~\cite{PRB, QuenchRev}.
        
        \subsection{Quantum Fisher information}
        
        The QFI is formally defined as~\cite{Liu_2019,helstrom1969,PETZ_2011}
        \begin{align}
        	\begin{aligned}\label{eq:def_qfi}
        		F_Q=\mathrm{Tr} \mathcal{L}_{\theta}^2 \rho_{\theta} ,
        	\end{aligned}
        \end{align}
        for a state described by a $\theta$-dependent  density operator $\rho_{\theta}$. Here $\mathcal{L}_{\theta}$ is the symmetric logarithmic derivative, defined by $\partial_{\theta} \rho_{\theta}=\left(\mathcal{L}_{\theta}\rho_{\theta}+\rho_{\theta}\mathcal{L}_{\theta}\right)/2$. 
        In many physical applications, the $\theta$ dependence of $\rho_{\theta}$ arises from a unitary transformation generated by a Hermitian operator $O$,
      \begin{align}\label{eq:unitary}
        	\rho_{\theta}=U_{\theta}^{\dagger}\rho U_{\theta},
        	\qquad
        	U_{\theta}=e^{i \theta O}.
       \end{align}
       In this case, using the spectral decomposition $\rho=\sum_{n}p_n \ket{n}\bra{n}$,  the QFI with respect to the generator $O$ can be expressed as~\cite{Liu_2019,Pezze-RMP}
             \begin{equation}
       	F_Q [\rho,O] = 2\sum_{nm}^{'}\frac{(p_n-p_m)^2}{p_n+p_m}
       	\lvert
       	\bra{n} O \ket{m}
       	\rvert^2
       	\leq 4 (\Delta O)^2,
       	\label{eq:def_qfi2}
       \end{equation}
       where the primed sum excludes terms with $p_n+p_m=0$, and
       $(\Delta O)^2=\mathrm{Tr} (\rho O^2)-\left[\mathrm{Tr}(\rho O)\right]^2$ is the variance of operator $O$ in the state $\rho$. The upper bound in this equation is always saturated by pure states. Specifically, for a pure state $\rho=\ket{\psi}\bra{\psi}$, the QFI  becomes~\cite{Liu_2019,Pezze-RMP}
      \begin{equation}
       	F_Q [\ket{\psi}, O] = 4\left(\bra{\psi}O^2\ket{\psi}-\bra{\psi}O\ket{\psi}^2\right).
       	\label{eq:def_qfi3}
       \end{equation}     
  For unitary parametrization Eq.~\ref{eq:unitary}, one can see from its definitions in Eqs.~\ref{eq:def_qfi2} and~\ref{eq:def_qfi3} that the QFI $F_Q$  has a strong dependence on the choice of generator $O$.
  
      The QFI admits  a geometric interpretation through its connection to the Bures distance~\cite{Caves1994,Liu_2019, Metrology,bengtsson2017}. The Bures distance between two states $\rho_1$ and $\rho_2$ is defined as \cite{nielsen2000}
      \begin{equation}
      	d_B (\rho_1,\rho_2) = \sqrt{2\left[1-f(\rho_1,\rho_2)\right]},
      \end{equation}
      where $f(\rho_1,\rho_2)$ is the fidelity
        \begin{equation}
      	f(\rho_1,\rho_2) = \mathrm{Tr}\left[\sqrt{\sqrt{\rho_1}\rho_2\sqrt{\rho_1}}\right].
      \end{equation}
	  The QFI is then related to the Bures distance between the two states $ \rho$ and $\rho_\theta = e^{-i\theta O} \rho e^{i\theta O}$ by~\cite{bengtsson2017,Liu_2019}
      \begin{equation}
      	F_Q[\rho,O] = \left[\lim_{\theta\rightarrow0}\frac{2d_B(\rho,e^{-i\theta O} \rho e^{i\theta O})}{|\theta|}\right]^2.
      \end{equation}
      It measures how rapidly the unitary evolution generated by $O$ moves the state $\rho_\theta = e^{-i\theta O} \rho e^{i\theta O}$ away from $\rho$, as quantified by the Bures distance.
      It can be viewed as the squared speed of the quantum state along the trajectory generated by the unitary transformation $\rho_\theta = e^{-i\theta O} \rho e^{i\theta O}$ in the Bures metric.

       The QFI was originally introduced in the context of quantum metrology for parameter estimation. For $\mathcal{N}$  repeated measurements, the QFI sets a lower bound on the variance of the unbiased estimation of the parameter $\theta$,
       \begin{equation}
       	(\Delta\theta)^2 \geq \frac{1}{\mathcal{N}F_Q} ,
       \end{equation}
       which is known as the quantum Cram\'er-Rao bound~\cite{helstrom1969,Caves1994,cramer1999,Pezze-RMP}. 
       This inequality relates the distinguishably of quantum states to the estimation of precision. In particular, the more sensitive the state $\rho$ is to a unitary transformation generated by operator $O$, the larger the QFI, and the more precisely the  parameter $\theta$ can be estimated.
       
       The QFI has also been widely used as a witness of multipartite entanglement~\cite{Hyllus_2012,toth2012,Pezze-RMP}. Consider, for example, a lattice model and a generator of the form $O=\sum_{i=1}^N O_i$, where $O_i$ is a local operator acting on site $i$. For a pure state, the QFI with respect to this generator can be expressed as~\cite{Si-2,Hauke_2016}
       \begin{align}
       	\begin{aligned}
       		F_Q[\rho, O]
       		=&4\sum_{i,j=1}^{N}\left[ \mathrm{Tr} (\rho O_i O_j)-\mathrm{Tr} (\rho O_i)\mathrm{Tr} (\rho O_j)\right]
       		\\
       		=&4\sum_{i\neq j} \mathrm{Tr} \left((\rho_{ij}-\rho_i\otimes\rho_j) O_i \otimes O_j\right)
       		\\
       		&+
       		4\sum_{i=1}^{N} \left[\mathrm{Tr} (\rho_i O_i^2 )- \left(\mathrm{Tr}(\rho_i O_i) \right)^2\right],
       	\end{aligned}
       	\end{align}
       where $\rho_i$ represents the reduced density matrix of site $i$ and $\rho_{ij}$ denotes the reduced density matrix for the pair of sites $i,j$. Each off-diagonal term corresponds to the covariance between sites $i\neq j$, and a nonzero result implies that the two-site reduced density matrix $\rho_{ij}$ deviates from the product state $\rho_i\otimes\rho_j$. It therefore probes the correlation between sites $i$ and $j$. The QFI sums these correlation across the entire system, and through its upper bounds, can serve as  a witness of the multipartite entanglement. Assuming that all local operators $O_i$ share the same eigenvalues, it has been shown that
       for $\mathcal{M}$-entangled states, the QFI is bounded by~\cite{Hyllus_2012,toth2012,Hauke_2016,Si-2}
       \begin{align}
       	F_Q \leq \mathcal{M} N (O_{max}-O_{min})^2,
       \end{align}
       where $O_{max}$ and $O_{min}$ are respectively the maximal and minimal eigenvalues of operator $O_i$. Using this bound, the QFI can be used to deduce the entanglement properties of the state. Specifically, a QFI violating this bound, or in other words satisfying $F_Q > \mathcal{M} N (O_{max}-O_{min})^2$, implies the presence of at least a $ \mathcal{M}+1$-partite entanglement.

		\subsection{Quenched $p+ip$ superfluid}

        \subsubsection{Model and dynamical phases}
		We consider a two-dimensional system of spinless fermions governed by the p-wave BCS Hamiltonian~\cite{Read}
		\begin{equation}
			H = \sum_\mathbf{k}\frac{k^2}{2m} c_{\mathbf{k}}^\dagger c_{\mathbf{k}}-\frac{2G}{m}\sum_{\mathbf{k},\mathbf{q}}{'}\mathbf{k}\cdot \mathbf{q} c_{\mathbf{q}}^\dagger c_{-\mathbf{q}}^\dagger c_{-\mathbf{k}}c_{\mathbf{k}}.
			\label{eq:BCS_Hamiltonian}
		\end{equation}
		Here $m$ is the mass of the fermions, $G>0$ is the coupling strength, and $c_\mathbf{k}$, $c_\mathbf{k}^\dagger$ are the fermionic annihilation and creation operators in momentum space, respectively. The primed sum restricts the momentum summation to the half plane with $k_y\geq 0$.  Using Anderson pseudospin representation~\cite{Anderson}, which offers an intuitive picture of the pairing structure, this Hamiltonian can be rewritten as
		\begin{equation}\label{eq:H}
			H = \sum_\mathbf{k}{'}\frac{k^2}{m}s_\mathbf{k}^z - \frac{2G}{m}\sum_{\mathbf{k},\mathbf{q}}{'} \mathbf{k}\cdot \mathbf{q} s^+_\mathbf{k}s_\mathbf{q}^-,
		\end{equation}
        where
		\begin{equation}
			\begin{aligned}
			&s_\mathbf{k}^z = \frac{1}{2}\left(c_\mathbf{k}^\dagger c_{\mathbf{k}}+c_{-\mathbf{k}}^\dagger c_{-\mathbf{k}}-1\right),
	\\
			&s_\mathbf{k}^+ = s_{\mathbf{k}}^x + i s_{\mathbf{k}}^y 
            = c_\mathbf{k}^\dagger c_{-\mathbf{k}}^\dagger,
		\quad
			 s_\mathbf{k}^- = s_{\mathbf{k}}^x - i s_{\mathbf{k}}^y = c_{-\mathbf{k}}c_\mathbf{k}.
			\end{aligned}
		\end{equation}
        In this description, $s_{\mathbf{k}}^z=1/2$ ($s_{\mathbf{k}}^z=-1/2$) indicates the presence (absence) of a Cooper pair $\left\lbrace \kb,-\kb \right\rbrace$, while $s_{\kb}^{+}$ ($s_{\kb}^{-}$) acts as the creation (annihilation) operator of such Cooper pairs. 
        
      This Hamiltonian admits two degenerate chiral ground states, corresponding to $p_x+ip_y$ and $p_x-ip_y$ pairing. In this work, initially, we prepare the system in the $p_x-ip_y$ ground state of the pre-quench Hamiltonian with coupling constant $G=G_i$, whose order parameter is
		 \begin{equation}
		 	\Delta_\mathbf{k} \equiv -\frac{2G_i}{m} \sum_{\mathbf{q}}{'} \mathbf{k}\cdot\mathbf{q} \langle s_{\mathbf{q}}^{-}\rangle = \Delta_0^{(i)}(k_x-ik_y).
		 	\label{eq:def_orderparam_pseudospins}
		 \end{equation}
       At time $t=0$, we then perform an instantaneous quantum quench, suddenly changing the coupling constant from $G_i$ to $G_f$~\cite{PRB}.
      After the quench, the nonequilibrium dynamics is governed by the effective post-quench Hamiltonian restricted to $p_x-ip_y$ sector
    	\begin{equation}
			H = \sum_\mathbf{k}{'}\frac{k^2}{m}s_\mathbf{k}^z - \frac{G_f}{m}\sum_{\mathbf{k},\mathbf{q}}{'} k q 
            e^{-i\phi_{\kb}+i\phi_{\mathbf{q}}}
            s^+_\mathbf{k}s_\mathbf{q}^-,
		\end{equation}
        where $\phi_{\kb}$ is the polar coordinate for momentum $\kb$.
     For convenience, we absorb this phase factor through the transformation  $\vex{s}_{\kb}^- \to \vex{s}_{\kb}^- e^{-i\phi_k}$. 

    In the thermodynamic limit, the dynamics of the Anderson pseudospins can be solved exactly using the self-consistent mean-field theory~\cite{PRB,yuzbashyan2005,BCS-MF}, and the corresponding equation of motion takes the form
		\begin{equation}\label{eq:EOM}
			\frac{d}{dt}\braket{\vex{s}_{\vex{k}}} = -\vex{B}_{\vex{k}} \times \braket{\vex{s}_{\vex{k}}}.
		\end{equation}
        Here the effective magnetic field is defined as
		\begin{equation}
			\vex{B}_{\vex{k}} \equiv-\frac{k^2}{m}\hat{z}-2k(\Delta_x\hat{x}+\Delta_y\hat{y}),
		\end{equation}
        and it couples the spins at all momenta through the self-consistent order parameter
		\begin{equation}
			\Delta(t) =\Delta_x-i\Delta_y=-\frac{G_f}{m}\sum_{\vex{k}}' k \left\langle s_{\vex{k}}^-(t)\right\rangle.
		\end{equation}
        
\begin{figure}[tbp]
    \centering
    \includegraphics[width=0.9\columnwidth]{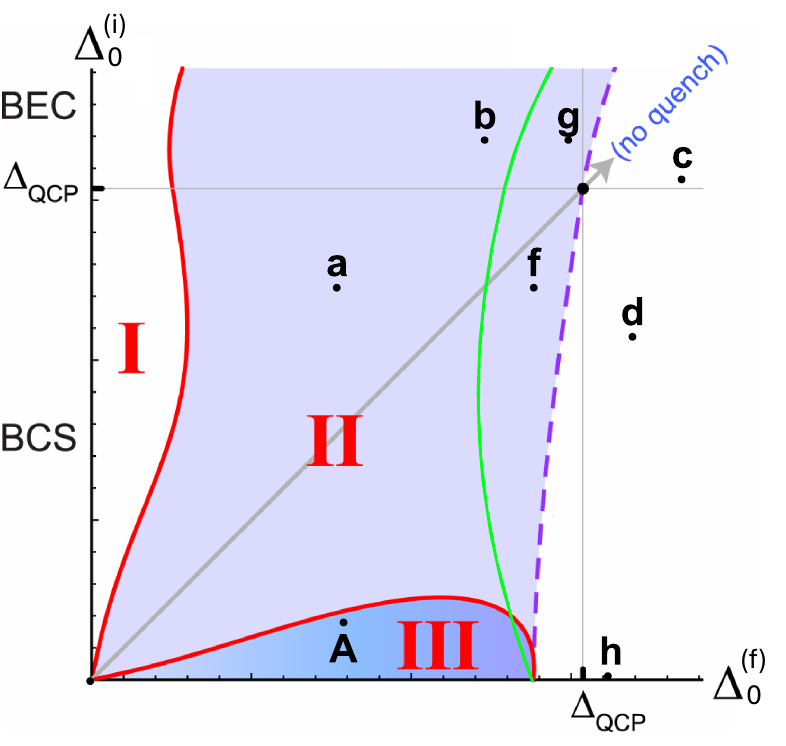}
    \caption{The quenched $p+ip$ superfluid phase diagram, from Ref.~\cite{PRB}, showing all the quenches considered in this paper. The horizontal and vertical axes show the order parameters $\Delta_0^{(f)}$ and $\Delta_0^{(i)}$ of the ground states of the post-quench and pre-quench Hamiltonians, respectively. The red lines separate three dynamical phases I, II and III, where the long-time asymptotic order parameter $\Delta_{\infty} (t)$ respectively decays to zero, approaches a nonzero constant, and oscillates periodically. The dashed purple line, where $\mu_{\infty}=0$, extends the equilibrium topological quantum critical point $\Delta_{\mathrm{QCP}}$ into the nonequilibrium regime. It separates the topological region with Green's function winding number $W=1$ ($\mu_{\infty}>0$), which hosts Majorana edge states, on its left, from the trivial region with $W=0$ ($\mu_{\infty}<0$), which does not host edge states, on its right.
    The green line marks the condition $m\Delta_\infty^2 = \mu_\infty$. Quenches to its right have $m\Delta_\infty^2 > \mu_\infty$, while those to its left have $m\Delta_\infty^2 < \mu_\infty$.
    Quenches in phase II (III) are labelled with lower (upper) case letters. 
    Throughout the paper, we have set the mass $m=1$ and particle density $n=0.825$.}
    \label{fig:phase_diag}
\end{figure}

	This equation of motion can be solved exactly using
    the Lax construction~\cite{PRB,yuzbashyan2005}. It has been found that, depending on the quench coordinate $(G_f, G_i)$, the quenched $p$-wave superfluid can fall into one of three dynamical phases, distinguished by the behaviour of the order parameter $\Delta(t)$ in the long time asymptotic limit~\cite{PRB,PRL,Levitov,Yuzbashyan2006}.
        In all three phases, the long-time asymptotic order parameter can be expressed as 
        \begin{equation}\label{eq:Deltainf}
			\Delta(t)=\Delta_{\infty}(t)e^{-2i\mu_{\infty}t},
		\end{equation}
        where the real constant $\mu_{\infty}$ is the asymptotic chemical potential.
        In phase I, $\Delta_{\infty}(t)$  vanishes asymptotically, $\Delta_{\infty}(t)\rightarrow 0$, and therefore $\mu_{\infty}$ no longer has a well defined meaning. So we set it to $\mu_{\infty}=0$ in Eq.~\ref{eq:Deltainf}. In phase II, $\Delta(t)$ approaches a nonzero constant, $\Delta_{\infty}(t) \rightarrow\Delta_\infty$. In phase III, $\Delta(t)$  exhibits persistent oscillation with period $T$, $ \Delta_\infty(t) = \Delta_\infty(t+T)$, and the system becomes a Floquet system generated by its own dynamics~\cite{PRL}.
       Fig.~\ref{fig:phase_diag} shows the quench phase diagram, adapted from~\cite{PRB}, with these three distinct dynamical phases. There, instead of $(G_f, G_i)$, each quench is represented by a single point in the phase diagram is labelled by the pairing amplitudes associated with the ground state of the pre-quench and post-quench Hamiltonians $(\Delta_0^{(f)},\Delta_0^{(i)})$. These nonequilibrium phases are separated by the red lines.

\subsubsection{Pseudospin dynamics}

        We now examine the pseudospin texture of the long time asymptotic states in these three dynamical phases and compare them with the equilibrium ground state. In equilibrium, 
        the ground-state pseudospin $\vex{s}_{\kb}$ aligns parallel to the effective  field $\vex{B}_{\kb}+2\mu_0 \hat{z}$, 
        \begin{equation}\label{eq:s0}
        \begin{aligned}
            &\langle \vex{s}_\mathbf{k}\rangle
            =
            -\frac{\Delta_0 k}{2E_{\kb}^0}\hat{x}
            -\frac{\xi_\mathbf{k}^0}{2E_{\kb}^0}\hat{z},
            %
            \end{aligned}
            \end{equation}
        Here $\xi_\mathbf{k}^0=k^2/2m-\mu_0$ is the single particle energy measured relative to the chemical potential $\mu_0$, and $E_{\kb}^0=E(k;\Delta_0,\mu_0)$ is the quasiparticle energy 
        \begin{align}
            E(k;\Delta_0,\mu_0)=\sqrt{(\frac{k^2}{2m}-\mu_0)^2+\Delta_0^2 k^2}.
        \end{align}
        
        Following the instantaneous quench of the coupling constant, the pseudospin $\vex{s}_{\kb}$ now precesses around the effective magnetic field $\vex{B}_{\kb}$, which shifts to $\vex{B}_{\kb} \to \vex{B}_{\kb} + 2\mu_{\infty} \hat{z}$ in the rotating frame~\cite{PRB}
		\begin{equation}\label{eq:rot}
			s^{-}_{\vex{k}}(t) \to s^{-}_{\vex{k}}(t)e^{i2\mu_{\infty}t}.
		\end{equation}
        In phase I, the asymptotic order parameter $\Delta_{\infty}(t)$ vanishes, so that the magnetic field $\vex{B}_{\kb}$ points in the $\hat{z}$ direction. The pseudospin then precesses around $\hat{z}$ axis as~\cite{PRB}
        \begin{equation}\label{eq:S-I}
        \begin{aligned}
             \langle \vex{s}_\mathbf{k} \rangle
             =&
             \frac{\sqrt{1-\gamma_\mathbf{k}^2}}{2}\left[\cos{(\frac{k^2}{m} t)\hat{x}+\sin{(\frac{k^2}{m}t)}\hat{y}}\right]
             +\frac{\gamma_\mathbf{k}}{2}\hat{z},
        \end{aligned}     
        \end{equation}
        Here $\gamma_\mathbf{k}$ is the Cooper-pair distribution function, related to the occupation number of the ground (excited) Cooper pair (see discussion below)~\cite{PRB,PRA}. In phase I and II, it equals two times the projection of $\mathbf{s}_\mathbf{k}$ onto $-\hat{\mathbf{B}}_\mathbf{k}=-\frac{{\mathbf{B}}_\mathbf{k}}{|{\mathbf{B}}_\mathbf{k}|}$. Its explicit expression can be determined from the conservation of the Lax vector norm~\cite{PRB}.

        In phase II, the asymptotic order parameter $\Delta_{\infty}$ assumes a nonzero constant, and the pseudospin $\vex{s}_{\kb}$ now precesses around an effective field $\hat{\vex{B}}_{\vex{k}}$ that is momentum dependent~\cite{PRB}:
        \begin{align}\label{eq:S-II}
        \begin{aligned}
        \langle \vex{s}_\mathbf{k} \rangle = &\frac{\sqrt{1-\gamma_\mathbf{k}^2}}{2}\left[\cos{(2E_\mathbf{k}^{\infty}t)\hat{\mathbf{B}}_\mathbf{k}\times\hat{y}+\sin{(2E_\mathbf{k}^{\infty}t)\hat{y}} }\right]\\& -\frac{\gamma_\mathbf{k}}{2}\hat{\mathbf{B}}_\mathbf{k},
        \end{aligned}
        \end{align}
        with frequency given by two times the quasiparticle energy $E_\mathbf{k}^{\infty}=E(k;\Delta_{\infty},\mu_{\infty})$. Note that Eq.~\ref{eq:S-II} reduces to its phase I counterpart Eq.~\ref{eq:S-I} as $\Delta_{\infty},\mu_{\infty} \to 0$.
        
		Lastly in phase III, the asymptotic order parameter and as a result  the effective field $\hat{\vex{B}}_{\vex{k}}$ becomes time periodic. 
		The pseudospin now acquires a form~\cite{PRA}
        \begin{align}\label{eq:s-III}
        \begin{aligned}
             \langle \vex{s}_\mathbf{k} \rangle 
             = &             
             \sqrt{\frac{1-\gamma_{\kb}^2}{2}} \left( 
             \vex{s}_{1,\mathbf{k}}(t)e^{-2iE_{\kb}^{(F)}t}
             +
             \vex{s}_{2,\mathbf{k}}(t)e^{2iE_{\kb}^{(F)}t}
             \right)
             \\
             &-\gamma_{\kb} \vex{s}_{0,\mathbf{k}}(t),
        \end{aligned}
        \end{align}
        where $E_{\kb}^{(F)}$ is the Floquet quasiparticle energy defined modulo the oscillation frequency $\Omega$. $\vex{s}_{a,\mathbf{k}}(t)$, $a=0,1,2$, are all time periodic with period $T$, and they satisfy 
        \begin{align}
        \vex{s}_{1,\kb}=\vex{s}_{2,\kb}^*,\quad
        |s_{0,\kb}|=|s_{1,\kb}|=\frac{1}{2},\quad
        \vex{s}_{0,\kb}\cdot\vex{s}_{1,\kb}=0.
        \end{align}
        The explicit expressions for $E_{\kb}^{(F)}$ and $\vex{s}_{a,\mathbf{k}}(t)$, $a=0,1,2$, are given in Ref.~\cite{PRA}.
        
        
		\subsubsection{Coherence factors}
        
		Apart from the pseudospin, another convenient way of describing the system is the coherence-factor representation. The many-body wave function $|\Psi\rangle$ of the system assumes a BCS product form both in and out of equilibrium~\cite{PRB,Levitov,PRA},
         \begin{equation}\label{eq:WF-BCS}
			|\Psi_{\msf{BCS}}(t)\rangle = \prod_\mathbf{k}{'}\left[u_\mathbf{k}(t) + v_\mathbf{k}(t)c_\mathbf{k}^\dagger c_\mathbf{-k}^\dagger\right]|0\rangle.
		\end{equation}
        These coherence factors are related to the pseudospin via
        \begin{align}
        \begin{aligned}
            \braket{s_{\kb}^-}=u_{\kb}^*(t)v_{\kb}(t),
            \qquad
            \braket{s_{\kb}^z}=
            \frac{1}{2}\left(|v_{\kb}(t)|^2-|u_{\kb}(t)|^2\right).
        \end{aligned}
        \end{align}
        In equilibrium, the coherence factors are time independent and assume the form~\cite{Read}
        \begin{align} \label{eq:uv0}
        \begin{aligned}
        &u_\mathbf{k}(t)=
        u_\mathbf{k}^0 = \sqrt{\frac{1}{2}\left(1+\frac{\xi_\mathbf{k}^0}{E_\mathbf{k}^0}\right)},
        \\
        &v_\mathbf{k}(t)=
        v_\mathbf{k}^0 = 
        -
        \sqrt{\frac{1}{2}\left(1-\frac{\xi_\mathbf{k}^0}{E_\mathbf{k}^0}\right)}.
        \end{aligned}
        \end{align} 
        
        For the post-quench state, the coherence factors $u_\mathbf{k}(t)$ and $v_\mathbf{k}(t)$ are solutions to the time-dependent Bogoliubov–de Gennes (BdG) equation
            \begin{equation}
		          i\frac{d}{dt}\begin{bmatrix}
				u_\mathbf{k}(t)\\v_\mathbf{k}(t)
			\end{bmatrix}=\begin{bmatrix}
				-\frac{k^2}{2m}+\mu_\infty & k\Delta_\infty(t)\\
				k\Delta_\infty(t) &\frac{k^2}{2m}-\mu_\infty
			\end{bmatrix}\begin{bmatrix}
				u_\mathbf{k}(t)\\v_\mathbf{k}(t)
			\end{bmatrix}.  
		\end{equation}
		They can be expressed  as the superposition~\cite{PRB,PRA}
		\begin{align}\label{eq:uv}	
        \begin{aligned}
			\begin{bmatrix}
				u_\mathbf{k}(t)\\
				v_\mathbf{k}(t) 
			\end{bmatrix} 
            =
            &
            \sqrt{\frac{1-\gamma_\mathbf{k}}{2}}\begin{bmatrix}
				\tilde{u}_\mathbf{k}(t)\\\tilde{v}_\mathbf{k}(t)
			\end{bmatrix}e^{iE_\mathbf{k}t} 
            \\&
            +\sqrt{\frac{1+\gamma_\mathbf{k}}{2}}
            \begin{bmatrix}
				-\tilde{v}_\mathbf{k}^*(t)\\\tilde{u}_\mathbf{k}^*(t)
			\end{bmatrix}e^{-iE_\mathbf{k}t+i\Gamma_k}.
     \end{aligned}       
    \end{align}
    The first term here is referred to as the ground state and the second term the excited state, and both of them solve the BdG equation. Their relative weight is determined by the Cooper-pair distribution function $\gamma_{\vex{k}}$. Specifically, $\frac{1}{2} (1 \mp \gamma_{\kb})$ gives the occupation number of ground (excited) Cooper pair at momentum $\kb$. 
    The time reversal symmetry of the equation of motion restricts the relative phase between the amplitudes of these two solutions to $\Gamma_k=0$ or $\pi$~\cite{PRA,PRB}. Throughout this paper, we set $\Gamma_k=0$ in phase II following the explicit phase II pseudospin solution in Ref.~\cite{PRB}, and also in phase III by continuity across the II-III boundary
    \footnote{Changing $\Gamma_k$ from $0$ to $\pi$ reverses the sign of certain oscillatory contributions to the QFI (specifically, $f^{(2)}_{\kb}$ in Eq.~\ref{eq:f-II} and $f^{(2)}_{n,\kb}$ in Eq.~\ref{eq:f-iii}),
    but it does not affect the main qualitative properties of the QFI discussed below. In particular, it does not affect the characteristic features of the QFI in three dynamical phases as well as its ability to distinguish between the topological and trivial initial states discussed below for phase II quenches, although it reverses the sign of the QFI spectral weight associated with the topological and trivial cases.}.
    

    
     In phases I and II, the ground and excited states coherence factors $\tilde{u}_\mathbf{k}(t)={u}_\mathbf{k}^{\infty}$ and $\tilde{v}_\mathbf{k}(t)={v}_\mathbf{k}^{\infty}$ are time independent and take the same form as their equilibrium counterparts in Eq.~\ref{eq:uv0}, but with $\xi_{\kb}^0$ and $E_{\kb}^0$ replaced by $\xi_{\kb}^\infty=\frac{k^2}{2m}-\mu_{\infty}$ and $E_{\kb}^\infty=E(k;\Delta_{\infty},\mu_{\infty})$. Particularly, in phase I, $\Delta_{\infty}=0$, so $E_{\kb}^{\infty}=|\xi_{\kb}^{\infty}|$, and the coherence factors reduce to ${u}_\mathbf{k}^{\infty}=1$ and ${v}_\mathbf{k}^{\infty}=0$ for $\xi_k^{\infty}>0$, and ${u}_\mathbf{k}^{\infty}=0$ and ${v}_\mathbf{k}^{\infty}=-1$ for $\xi_k^{\infty}<0$.
    In phase III, with time periodic order parameter $\Delta_{\infty}(t)$, the two solutions become the Floquet states, and 
    $\tilde{u}_\mathbf{k}(t)$ and $\tilde{v}_\mathbf{k}(t)$ are also time periodic with the same period. The corresponding energy $E_{\kb}=E_{\kb}^{(F)}$ is the Floquet quasienergy, defined modulo $\Omega=2\pi/T$. For their explicit expressions, we refer the reader to Ref.~\cite{PRA}.

    \subsubsection{Topological characterization}
    
    In equilibrium, the system can undergo a topological quantum phase transition between the topologically nontrivial BCS phase ($\Delta_0<\Delta_{\rm{QCP}}$) and the  topologically trivial BEC phase ($\Delta_0>\Delta_{\rm{QCP}}$) as the pairing strength is varied, where $\Delta_{\rm{QCP}}$ denotes the order parameter amplitude at the quantum critical point~\cite{Read}. The topology of the ground state can be characterized by the pseudospin winding number $Q$, which measures the winding of the Anderson pseudospin texture~\cite{Read,volovik,PRB},
    \begin{align}
    \begin{aligned}\label{eq:Q}
    Q
    =\frac{1}{4\pi}
    \int dk_x dk_y \hat{\vex{s}}_{\vex{k}}
    \cdot 
    \left(
    \frac{\partial \hat{\vex{s}}_{\vex{k}}}{\partial k_x} \times
    \frac{\partial\hat{\vex{s}}_{\vex{k}}}{\partial k_y} 
    \right),
    \end{aligned}
    \end{align}
    where $\hat{\vex{s}}_{\kb}={\vex{s}_{\kb}}/|{\vex{s}_{\kb}}|$.
    Because of the chiral $p+ip$ pairing symmetry, the Anderson pseudospins at fixed momentum magnitude $k$ wind with the polar angle $\phi_{\kb}$, so that the pseudospin winding number reduces to $Q=\frac{1}{2}(\hat{s}_{0}^z-\hat{s}_{\infty}^z)$~\cite{PRB}. Note that the pseudospin $\vex{s}_{\kb}$ points along the $-\hat{z}$ direction at large momentum $\kb$, $\vex{s}_{\infty}=-\frac{1}{2}\hat{z}$, while its direction at $\kb=0$ is determined by the sign of the chemical potential, $\vex{s}_{0}=\frac{1}{2}\sgn(\mu_0)\hat{z}$. One therefore finds $Q=1$ in the topological BCS phase with $\mu_0>0$ and $Q=0$ in the topologically trivial BEC phase with $\mu_0<0$. For a system with a boundary, the pseudospin winding number $Q$ predicts the presence ($Q=1$) or absence ($Q=0$) of Majorana edge modes~\cite{Read,Alicea_2012,PRB}.
        
    Following a  quantum quench, however, this correspondence between the pseudospin winding number and Majorana edge modes breaks down. The equation of motion shows that the pseudospin $\vex{s}_{\kb}$ at zero momentum $\kb=0$ is frozen in time, since its effective field vanishes $B_{\kb=0}=0$. As a result, $Q$ remains conserved across the quench and reflects the topology of the pre-quench state~\cite{PRB}. 
    The topology of asymptotic BdG Hamiltonian in phase II is instead characterized by a different winding number $W$, originally introduced in terms of the retarded single-particle Green's function~\cite{PRB,volovik}. It can also be expressed
   in the same form as $Q$ in Eq.~\ref{eq:Q}, with $\hat{\vex{s}}_{\kb}$ replaced by the unit effective field $\hat{\vex{B}}_{\kb}$ in the rotating frame (Eq.~\ref{eq:rot}), and reduces to $W=\frac{1}{2} (\hat{B}_0^z-\hat{B}_{\infty}^z)$~\cite{PRXQuantum}. As a result, $W$ is determined by the sign of the asymptotic chemical potential $\mu_{\infty}$, taking the values of $W=1$  in the region with positive asymptotic chemical potential $\mu_{\infty}>0$, and $W=0$ for negative $\mu_{\infty}$. It is the Green's function winding number $W$, rather than the pseudospin winding number $Q$ that determines whether the asymptotic quasiparticle spectrum supports Majorana edge modes or not~\cite{PRB}.
   In particular, there exist quenches with $W\neq Q$, for which the Majorana edge modes appear or disappear in the  quasiparticle spectrum  after the quench~\cite{PRB}.
   In the phase diagram~\cite{PRB} shown in Fig.~\ref{fig:phase_diag}, the $\mu_{\infty}=0$ line (dashed purple line) indicates the nonequilibrium extension of the topological quantum phase transition.  To its left (right) in phase II, we have  $\mu_{\infty}>0$ ($\mu_{\infty}<0$) and $W=1$  ($W=0$), and the system supports (does not support) Majorana edge states.
   
   The above discussion applies to the dynamical phase II with constant order parameter. In phase I, the order parameter decays to zero and the post-quench state is gapless. While the conserved winding number $Q$ characterize the pseudospin textures associated with the initial state, $W$ is ill defined in this gapless phase~\cite{PRB}. By contrast, in phase III, where the order parameter exhibits persistent oscillation, the resulting Floquet quasiparticle spectrum supports Majorana edge modes for all quenches within this phase~\cite{PRL}.

\section{Results: QFI in and out of equilibrium}~\label{sec:results}

In this work, we study the QFI of the quenched $p+ip$ superfluid in the post-quench long time asymptotic BCS state $\ket{\Psi_{\msf{BCS}}(t)}$ (Eq.~\ref{eq:WF-BCS}), using  the particle number in a subsystem $A$ 
\begin{align}
    N_A=\sum_{\vex{r}\in A} c_{\mathbf{r}}^{\dagger}c_{\mathbf{r}},
\end{align}
as the generator.
This QFI measures the distinguishability between the states $\ket{\Psi_{\msf{BCS}}(t)}$ and $e^{-i\theta N_A}\ket{\Psi_{\msf{BCS}}(t)}$ for an infinitesimally small parameter $\theta$.
The subsystem is taken to be large but subextensive, with linear size $L_A$ much smaller compared with that of the total system $L_A \ll L$. 
In this regime, the corresponding QFI can be evaluated within the self-consistent mean field description~\cite{BCS-MF,PRB}, using the pseudospin or equivalently the coherence factors dynamics reviewed in Sec.~\ref{sec:background}.  
The relative correction to the QFI beyond mean field is expected to be of the order of $O(V_A/V)$ and therefore negligible in the limit $V_A/V\to 0$, with $V_A$ ($V$) being the volume of the subsystem $A$ (total system).
See Ref.~\cite{BCS-MF} for a detailed discussion of the applicability of the mean field description for local and global observables in BCS dynamics.
In contrast, the total particle number $N=\sum_{\vex{r}} c_{\mathbf{r}}^{\dagger}c_{\mathbf{r}}$ commutes with the Hamiltonian $H$ in Eq.~\ref{eq:H}, $[H,N]=0$, and as a result the total particle number QFI in the exact many-body post-quench state evolved under $H$ remains invariant in time. The QFI evaluated in the BCS mean-field state, $F_Q[{N}, |\Psi_{\msf{BCS}}(t)\rangle]$, on the other hand, is time dependent, but its time dependence is an artifact of the mean field approach and does not represent the exact many body dynamics of the post quench system.

Using the pure-state expression for the QFI in Eq.~\ref{eq:def_qfi3} together with the BCS form of the many-body wavefunction in Eq.~\ref{eq:WF-BCS}, one finds the QFI associated with $N_A$ for the BCS product state $\ket{\Psi_{\msf{BCS}}(t)}$ 
can be expressed as
\begin{align}\label{eq:FQA}
\begin{aligned}
    &F_Q[{N}_A, |\Psi_{\msf{BCS}}(t)\rangle]
    =
    4\sum_{\kb,\kb'}
    |g(\kb-\kb')|^2
    \\
    &\times
    \left[
    |v_{\kb}(t)|^2|u_{\kb'}(t)|^2+
    \re
    \left(
    u_{\kb}(t)v^*_{\kb}(t)u_{\kb'}^*(t)v_{\kb'}(t)e^{i\phi_{\kb}-i\phi_{\kb'}}
    \right)
    \right],
\end{aligned}
\end{align}
where
\begin{align}
    \begin{aligned}
    g(\kb)= \frac{1}{V}\sum_{\vex{r} \in A} e^{-i\kb\cdot \vex{r}}.
    \end{aligned}
\end{align}
Equivalently, the QFI can be rewritten in terms of the Anderson pseudospins as
\begin{align}\label{eq:FQA2}
\begin{aligned}
    &F_Q[{N}_A, |\Psi_{\msf{BCS}}(t)\rangle]
    =
    4\sum_{\kb,\kb'}
    |g(\kb-\kb')|^2
    \\
    &\times
    \left[
    \frac{1}{4}-\braket{s_{\kb}^z(t)}\braket{s_{\kb'}^z(t)}+
    \re
    \left(
    \braket{s_{\kb}^+(t)}\braket{s_{\kb'}^-(t)}e^{i\phi_{\kb}-i\phi_{\kb'}}
    \right)
    \right].
\end{aligned}
\end{align}
In both the equations above, the pseudospins and coherence factors are defined in the rotated frame, related to the original ones by the transformation ${s}_{\kb}^- \to {s}_{\kb}^- e^{-i\phi_{\kb}}$. Additionally, we note that both expressions are invariant under any momentum-independent phase rotation of $s^-_{\kb}$, such as the rotating-frame transformation in Eq.~\ref{eq:rot}. The results in Eqs.~\ref{eq:s0},~\ref{eq:S-I},~\ref{eq:S-II}, and~\ref{eq:s-III} (Eqs.~\ref{eq:uv0} and~\ref{eq:uv}) can therefore be directly substituted into Eq.~\ref{eq:FQA2} (Eq.~\ref{eq:FQA}) to obtain the QFI in equilibrium and in three dynamical phases.

In the limit $L_A\to L$ when the subsystem $A$ is the entire system, one has $g(\kb-\kb')\to \delta_{\kb,\kb'}$, and Eqs.~\ref{eq:FQA} and~\ref{eq:FQA2} reduce to the QFI associated with total particle number $N$ for the BCS product state~\cite{QFI-BCS}:
\begin{align} \label{eq:QFI-uv}
   F_Q[{N}, |\Psi_{\msf{BCS}}(t)\rangle] = 8\sum_\mathbf{k}|\langle s_\mathbf{k}^- (t)\rangle|^2
    =8\sum_\mathbf{k}|u_\mathbf{k}(t)|^2|v_\mathbf{k}(t)|^2.
\end{align}
This, as mentioned above, is not the exact QFI, whose evaluation requires consideration beyond mean field.

To simplify the calculation, we assume that the linear size of subsystem $A$ satisfies
$k_s^{-1} \ll L_A\ll L$, where $k_s$ stands for the characteristic momentum scale over which the pseudospin configuration (in the original frame before applying the transformation ${s}_{\kb}^- \to {s}_{\kb}^- e^{-i\phi_{\kb}}$) varies appreciably. Using the fact that the factor $|g(\kb-\kb')|^2$ is sharply peaked when $|\kb-\kb'| \lesssim L_A^{-1} \ll k_s$,  we may approximate $\braket{\vex{s}_{\kb'}}\approx \braket{\vex{s}_{\kb}}$ in Eqs.~\ref{eq:FQA} and~\ref{eq:FQA2} and perform the remaining $\kb'$ summation using $\sum_{\kb'}|g(\kb')|^2=\frac{V_A}{V}$, which leads to
\begin{align} \label{eq:QFIA3}
\begin{aligned}
   f_Q = \frac{F_Q[{N}_A, |\Psi_{\msf{BCS}}(t)\rangle]}{V_A}
   &\approx \frac{8}{V}
   \sum_\mathbf{k}|\langle s_\mathbf{k}^- (t)\rangle|^2
   \\
    &=  \frac{8}{V}
    \sum_\mathbf{k}|u_\mathbf{k}(t)|^2|v_\mathbf{k}(t)|^2.
\end{aligned}    
\end{align}
Here for convenience we have introduced the subsystem QFI density $f_Q=F_Q[N_A,|\Psi_{\msf{BCS}}(t)\rangle]/V_A$.
We note that this approximated density coincides with the QFI density for the total particle number $N$ in the BCS product state, defined as $F_Q[{N}, |\Psi_{\msf{BCS}}(t)\rangle]/V$.

In the following, we first use the approximated expression for $f_Q$ in Eq.~\ref{eq:QFIA3} to evaluate the QFI density of the equilibrium  ground state and the post-quench long-time asymptotic state, in the regime $k_s^{-1} \ll L_A\ll L$. Through concrete examples 
\footnote{We consider in this paper a set of representative quenches in phases II and III. Some of these quench considered here (phase II quenches ``a'', ``b'', ``f'' , ``c'' and ``d'', and phase III quench ``A'', in Fig.~\ref{fig:phase_diag}) were previously studied in Ref.~\cite{PRA} in the context of spectroscopic probes of nonequilibrium systems. We revisit these quenches, together with a few additional representative quenches,
to study their QFI.}, 
we show how the post-quench QFI spectrum distinguishes the three dynamical phases, and particularly how it encodes the topology of the pre-quench state in phase II. Later in Sec.~\ref{sec:FC}, we discuss the regime in which the condition $L_A\gg k_s^{-1}$ is not satisfied, and the full expression in Eqs.~\ref{eq:FQA2} or~\ref{eq:FQA} need to be used. We demonstrate that many of the main features identified using the approximated expression for $f_Q$ in Eq.~\ref{eq:QFIA3} survive even outside the regime $L_A\gg k_s^{-1}$.

\subsection{QFI in equilibrium}\label{sec:F0}

We first consider the QFI for the equilibrium ground state.
Substituting the equilibrium coherence factors in Eq.~\ref{eq:uv0} or the pseudospin configuration in Eq.~\ref{eq:s0}, yields the equilibrium QFI
\begin{equation}\label{eq:qfi0}
F_Q^0 =8\frac{V_A}{V}\sum_\mathbf{k}|u_\mathbf{k}^0|^2|v_\mathbf{k}^0|^2
= 2 \frac{V_A}{V}\sum_\mathbf{k}\frac{\Delta_0^2 k^2}{(E_\mathbf{k}^0)^2}.
\end{equation}
Converting the summation to integral leads to
\begin{equation}\label{eq:qfi0-1}
\begin{aligned}
    f_Q^0 &= \frac{2m^2\Delta_0^2}{\pi}\int_{-\mu_0}^{\Lambda}
    d\xi
    \frac{\xi+\mu_0}{\xi^2 +2m\Delta_0^2(\xi+\mu_0)}
    \\
    &\simeq 
    \frac{2m^2\Delta_0^2}{\pi}\ln{\Lambda}+\mathcal{O}(1),
  \end{aligned}  
\end{equation}
where $\Lambda$ is the ultraviolet (UV) energy cutoff.
The full expression for the equilibrium QFI density, including the $\mathcal{O}(1)$ term, is given in Appendix~\ref{app:eqQFI}.
Unlike its $s$-wave counterpart~\cite{QFI-BCS}, the equilibrium $p$-wave QFI density is logarithmically 
UV divergent, due to the extra factor of $k$ in the pairing potential $\Delta_{\kb}$. Additionally, $f_{Q}^{0}$ is continuous across the quantum critical point at $\mu_0=0$\footnote{Higher derivatives of $f_{Q}^{0}$ can exhibit nonanalytic behavior at the transition}, 
providing no direct signature of the topological quantum phase transition in equilibrium.

\subsection{QFI in phase I}\label{sec:FI}

In phase I, the asymptotic order parameter vanishes, so the pseudospin $\vex{s}_{\kb}$ precesses around the $\hat{z}$ axis at frequency $k^2/m$. The magnitude of transverse pseudospin component $|\braket{{s}_{\kb}^{-}}|$ is conserved and is determined entirely by the Cooper pair distribution $\gamma_{\kb}$. As a result, the QFI density is time-independent in  phase I,
\begin{equation}\label{eq:F-I}
    f^{(\mathrm{I})}_Q =  \frac{2}{V} \sum_\textbf{k}(1-\gamma_\textbf{k}^2).     
\end{equation}
Since the pseudospins remain aligned with $-\hat{z}$ at large momentum, the Cooper-pair distribution function - the projection of $\vex{s}_{\kb}$ onto $-\hat{\vex{B}}_{\kb}$ - approaches $\gamma_{\kb}\rightarrow -1$ as $k\to \infty$. 
Consequently, the integrand $1-\gamma_{\kb}^2\to 0$ as $k\to \infty$, and one can show that the QFI in phase I is UV convergent, in contrast to the equilibrium case.
Conservation of the zero-momentum pseudospin $\vex{s}_{0}$ under the equation of motion  implies that in phase I, $\gamma_{0}=\sgn(\mu_0^{(i)})$, so the sign of $\gamma_{0}$ directly encodes the pseudospin winding number $Q$ of the pre-quench state
\begin{align}
    \gamma_{0}
    =
    \begin{cases}
        +1, & Q=1 \,(\mu_0^{(i)}>0),
        \\
        -1, & Q=0 \,(\mu_0^{(i)}<0).
    \end{cases}
\end{align}
However, since $F^{(\mathrm{I})}_Q$ depends on the distribution function only through $\gamma_{\kb}^2$, the topological information of the pre-quench state is also not directly accessible from the QFI in phase I.

\subsection{QFI in phase II}\label{sec:FII}

\begin{figure}[tbp]
    \centering
    \includegraphics[width=0.9\columnwidth]{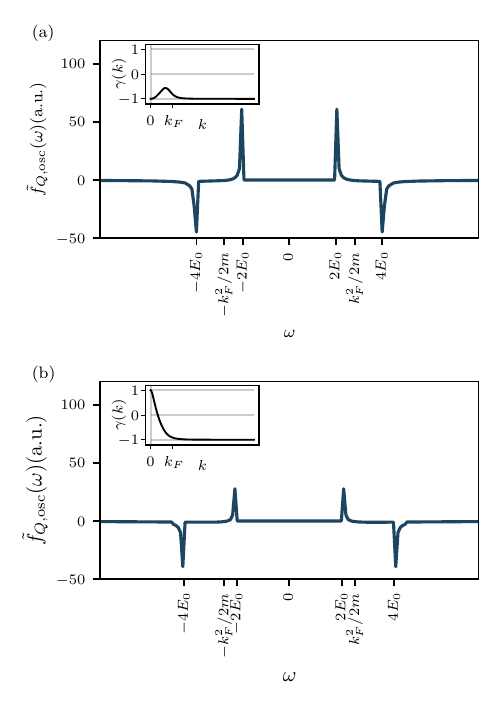}
    \caption{Fourier transform of the QFI density with the zero frequency spike omitted, i.e. the Fourier transform of the oscillating part of the QFI density $\tilde{f}_{Q,osc}(\omega)$, for  (a) the BCS-to-BCS quench ``a'', with coordinate $(\Delta_0^{(i)}, \Delta_0^{(f)})=(0.8,0.5)\Delta_\mathrm{QCP}$, and (b) the BEC-to-BCS quench ``b'', with $(\Delta_0^{(i)}, \Delta_0^{(f)})= (1.1,0.8)\Delta_\mathrm{QCP}$.  Both quenches lie to the left of the $\mu_{\infty}=m\Delta_{\infty}^2$ line in the phase diagram (i.e., satisfy $\mu_{\infty}>m\Delta_{\infty}^2$), and their QFI spectra exhibit van Hove singularities at $\omega=\pm 2E_0, \pm 4E_0$, with $E_0$ denoting the minimum quasiparticle energy (Eq.~\ref{eq:E0}). Insets show the corresponding Cooper-pair distribution function $\gamma(k)$. For quench ``a'' with $Q=W=1$, $\gamma(k)\to -1$ at both $k=0$ and $k\to\infty$, while for quench ``b'' with $Q=0$ and $W=1$, $\gamma(k)$ starts at $+1$ at $k=0$  and approaches $-1$ as $k\to\infty$.}
    \label{fig:II_ab}
\end{figure}

In phase II, the order parameter approaches a nonzero constant, and the pseudospin at momentum $\kb$ precesses around an effective field $\vex{B}_{\kb}$, whose direction depends on momentum $\kb$ and is no longer always aligned with $\hat{z}$. As a result, $|s_{\kb}^-|$ contains an oscillating component with the frequency set by the asymptotic quasiparticle energy $E_{\kb}^{\infty}$, making the QFI time dependent 
\begin{align}\label{eq:F-II}
        {f}_Q^{(\mathrm{II})}(t) =
        \frac{1}{V}
        \sum_\mathbf{k}
        \left[
        f_{ \mathbf{k}}^{(0)} 
        +
        f_{ \mathbf{k}}^{(2)} \cos(2 E_{\kb}^{\infty}t)
        +
        f_{ \mathbf{k}}^{(4)} \cos(4 E_{\kb}^{\infty}t)
        \right],
\end{align}
where
\begin{align}
\begin{aligned}\label{eq:f-II}
        f_{\mathbf{k}}^{(0)} 
        =\,& 
        \frac{(1+\gamma_\mathbf{k}^2)\Delta_\infty^2 k^2+2(1-\gamma_\mathbf{k}^2)(\xi_\mathbf{k}^{\infty})^2}{(E_\mathbf{k}^{\infty})^2},
\\
        f_{\mathbf{k}}^{(2)}
        =\,&
        +
        4\gamma_\mathbf{k}\sqrt{1-\gamma_\mathbf{k}^2}
        \frac{\Delta_\infty k \xi_\mathbf{k}^{\infty}}{(E_\mathbf{k}^{\infty})^2},
        \\
        f_{\mathbf{k}}^{(4)} 
        =\,&
        -(1-\gamma_\mathbf{k}^2)\frac{\Delta_{\infty}^2k^2}{(E_\mathbf{k}^{\infty})^2}.
\end{aligned}
\end{align}
Note that the long-time average of ${F}_Q^{(\mathrm{II})}(t)$ retains only the contribution from the time-independent term $f^{(0)}_{\kb}$, while the two remaining oscillating terms weighted by $f^{(2)}_{\kb}$ and $f^{(4)}_{\kb}$ average to zero, i.e., $  \overline{f_Q^{(\mathrm{II})}} =
        \sum_\mathbf{k}
        f_{ \mathbf{k}}^{(0)}$.

Similar to the QFI in phase I,  $f^{(0)}_{\kb}$ and $f^{(4)}_{\kb}$ depend on the distribution function $\gamma_{\kb}$ only through $\gamma_{\kb}^2$, and as a result they do not directly reveal the underlying system topology. By contrast, the sign of $f^{(2)}_{\kb}$ is determined by the signs of $\gamma_{\kb}$ and $\xi_{\kb}^{\infty}$, both of which carry topological information at $\kb=0$. In ${F}_Q^{(\mathrm{II})}(t)$, however, this topological information is buried by the momentum summation.

This hidden topological information can be recovered by Fourier transforming the QFI, which separates the  contributions of pseudospins according to their oscillation frequencies, or equivalently, their asymptotic quasiparticle energies $E_{\kb}^{\infty}$.
More specifically, the Fourier component
\begin{align}\label{eq:tF-II}
        \tilde{f}_Q^{(\mathrm{II})}(\omega) =
        \frac{1}{V}
        \sum_{a =0,2,4}\sum_\mathbf{k}
        \pi f_{ \mathbf{k}}^{(a)} 
        \left[\delta(\omega-a E_\mathbf{k}^{\infty})
        +\delta(\omega+aE_\mathbf{k}^{\infty})\right].
\end{align}
at frequency $\omega$ receives contributions only from pseudospins with quasienergies satisfying $\omega=\pm aE_\mathbf{k}$, for $a=0,2,4$.
Consequently, $f^{(0)}_{\kb}$ produces a single zero-frequency spike in the QFI spectrum $\tilde{F}_{Q}^{(\mathrm{II})}(\omega)$, while $f^{(2)}_{\kb}$ and $f^{(4)}_{\kb}$ give rise to continua over the frequency ranges $|\omega|\geq 2E_0$ and $|\omega|\geq 4E_0$, respectively. Here $E_0$ denotes the minimum of the asymptotic quasiparticle energy spectrum $E_{\kb}^{\infty}$:
\begin{equation}\label{eq:E0}
E_0 = 
    \begin{cases}
       \sqrt{m}\Delta_\infty\sqrt{2\mu_\infty-m\Delta_\infty^2}, 
       & \mu_\infty > m\Delta_\infty^2,
      \\
      |\mu_\infty|,  
      &  \mu_\infty< m\Delta_\infty^2.
    \end{cases}
\end{equation}
Combining everything, the QFI spectrum $\tilde{F}_{Q}^{(\mathrm{II})}(\omega)$ consists of two continua, $(-\infty, -2E_0]$ and $[2E_0, \infty)$, as well as a single spike at $\omega=0$.

\begin{figure}[tbp]
    \centering
    \includegraphics[width=0.9\columnwidth]{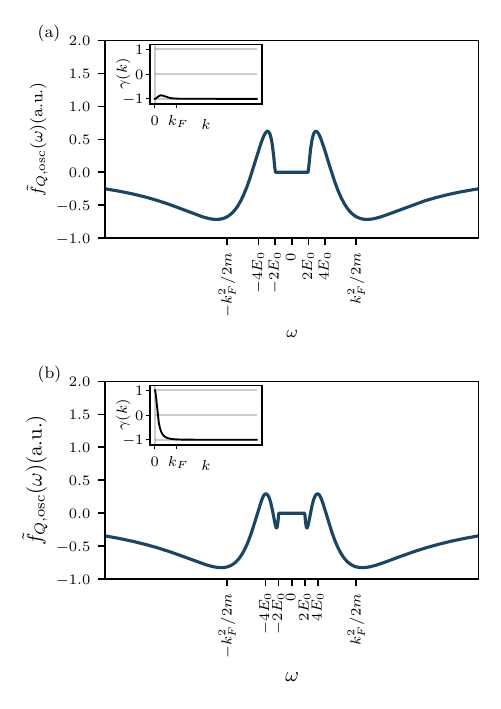}
    \caption{Same as Fig.~\ref{fig:II_ab}, but for (a) the BCS-to-BCS quench ``f'' with coordinates $(\Delta_0^{(i)}, \Delta_0^{(f)})=(0.8,0.9)\Delta_\mathrm{QCP}$ and (b) the BEC-to-BCS quench ``g'' with  $(\Delta_0^{(i)}, \Delta_0^{(f)}) = (1.1, 0.97)\Delta_\text{QCP}$. Both quenches satisfy $\mu_\infty<m\Delta_\infty^2$, and their QFI spectra as predicted vanishes exactly at the continuum edges $\omega = \pm 2E_0$.  
    As $\omega$  approaches $ \pm (2E_0+0^+)$, the sign of $\tilde{f}_{Q,\mathrm{osc}}(\omega)$ is positive (negative) for quench ``f'' (``g''), revealing the winding number $Q=1$ ($Q=0$) that characterizes the topology of pre-quench state. Insets show the corresponding Cooper-pair distribution function $\gamma_{\kb}$, which satisfies $\gamma_k=-1$ ($\gamma_k=+1$) at $k=0$ for quench ``f'' (``g'') with $W=Q$ ($W\neq Q$).
    }
    \label{fig:II_fg}
\end{figure}

When $\mu_\infty < m\Delta_\infty^2$, the minimum quasiparticle energy $E_0$ occurs at $k=0$. Since $f_{\kb}^{(2)}\propto k$ vanishes there, the QFI spectrum turns on continuously from zero as $\omega \to \pm (2E_0+0^{+})$. 
In contrast, when $ \mu_\infty > m\Delta_\infty^2$, the minimum quasiparticle energy $E_0$ occurs at finite momentum $k^*=\sqrt{2m(\mu_\infty-\Delta_\infty^2m)}$, where $f_{\kb}^{(2)}$ remains finite. At the edge of the continuum $\omega=\pm 2E_0$, only the pseudospins with $k=k^*$ contribute. The finite weight $f_{k^*}^{(2)}$, together with the divergent Jacobian $1/|dE^{\infty}/dk|$ at $k=k^*$, produces van Hove singularities at the continuum edges~\cite{PRB,PRA}. Similar van Hove singularities also appear at $\omega=\pm 4E_0$, due to the $f^{(4)}_{\kb}$ contribution. In conclusion, the behaviour of the QFI spectrum at the continuum edges -  whether it vanishes or exhibits van Hove singularities - directly determines the sign of $m\Delta_\infty^2 - \mu_\infty$.

In the case of $\mu_\infty < m\Delta_\infty^2$, when the QFI spectrum vanishes at the continuum edge $\omega=\pm 2 E_0$, the system topology is encoded in how it departs from zero at the edges, that is in its initial sign just inside the continua $\sgn \tilde{F}_{Q}^{(\mathrm{II})}(\pm (2 E_0+0^+))$. 
Near the continuum edge, the spectrum is dominated by small but finite momenta. In this regime, only $f^{(2)}_{\kb}$ contributes, whose sign is determined by $\lim_{\kb\to 0}\gamma_{\kb}\xi_{\kb}^{\infty}=-\sgn(\gamma_0\mu_{\infty})$ (Eq.~\ref{eq:f-II}) due to continuity.
In phase II, the conservation of zero-momentum pseudospin $\vex{s}_{0}$ implies that  the distribution function at zero momentum is given by $\gamma_{0}=-\sgn(\mu_0^{(i)}\mu_{\infty})$, namely
\begin{align}\label{eq:gamma_0_II}
    \gamma_{0}
    =
    \begin{cases}
        +1, & W \neq Q \,(\mu_0^{(i)}\mu_{\infty}<0),
        \\
        -1, & W=Q \,(\mu_0^{(i)}\mu_{\infty}>0).
    \end{cases}
\end{align}
Together with the fact that $\gamma_{\kb}\to -1$ as $k\to \infty$, this implies that the Cooper-pair distribution function has an odd number of zeros when $W\neq Q$ and an even number when $W=Q$~\cite{PRB}.
Combining these results yields $\sgn \tilde{F}_{Q}^{(\mathrm{II})}(\pm (2 E_0+0^+))=\sgn(\mu_{0}^{(i)})$, showing that 
the sign of the QFI spectrum immediately inside the continuum directly reveals the topology of the pre-quench state. It is positive for phase II quenches from the topological BCS phase ($\mu_0^{(i)}>0$), and negative for phase II quenches from the trivial BEC phase ($\mu_0^{(i)}<0$). This conclusion holds only when the QFI spectrum itself vanishes at the continuum edge, i.e., when $\mu_\infty< m\Delta_\infty^2$. When $ \mu_\infty> m\Delta_\infty^2$, the continuum edge is instead dominated by contributions from finite momentum $k^*$ that produce van Hove singularities. The edge behaviour therefore no longer provides a direct probe of either the pre-quench or post-quench topology.

\begin{figure}[tbp]
    \centering
    \includegraphics[width=0.9\columnwidth]{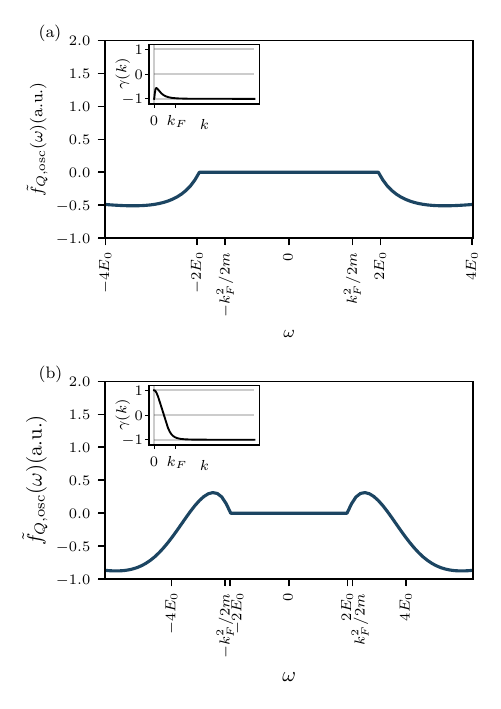}
    \caption{
    Same as Fig.~\ref{fig:II_ab}, but for (a) the BEC-to-BEC quench ``c'' with coordinates $(\Delta_0^{(i)}, \Delta_0^{(f)})=(1.02,1.2)\Delta_\mathrm{QCP}$ and (b) the BCS-to-BEC quench ``d'' with  $(\Delta_0^{(i)}, \Delta_0^{(f)}) = (0.7,1.1)\Delta_\text{QCP}$. Both quenches satisfy the condition $\mu_\infty<m\Delta_\infty^2$, and as a result the QFI spectrum vanishes at the continuum edges.
    Just inside the continua near the edge, the sign of $\tilde{f}_{Q,\mathrm{osc}}(\omega)$ is negative (positive) for quench ``c'' (``d''), corresponding to a trivial (nontrivial) initial state with $Q=0$ ($Q=1$). Both quenches share the same $W=0$, and since $W=Q$ ($W\neq Q$) for quench ``c'' (``d''), the distribution function $\gamma$ shown in the inset exhibits an even (odd) number of zeros.
     }
    \label{fig:II_cd}
\end{figure}

\begin{figure}[tbp]
    \centering
    \includegraphics[width=0.9\columnwidth]{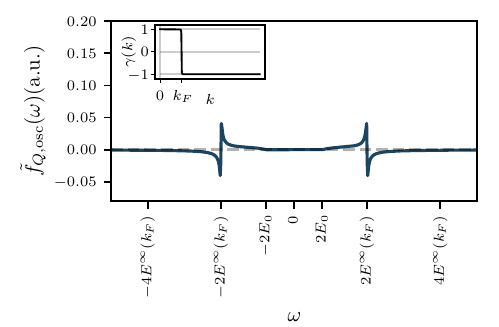}
    \caption{
        Same as Fig.~\ref{fig:II_ab}, but for the BCS-to-BEC quench ``h'' with  $(\Delta_0^{(i)}, \Delta_0^{(f)}) = (0.01, 1.05)\Delta_\text{QCP}$. The Cooper-pair distribution function shown in the inset becomes approximately a step function with the step located at the Fermi momentum $k_F$. As a result, the QFI spectrum exhibits pronounced peaks near $\omega=\pm2E^\infty(k_F)$. 
    }
    \label{fig:II_hl}
\end{figure}

The QFI spectra for representative phase II quenches confirm the analytical predictions derived above. 
In Figs.~\ref{fig:II_ab}-\ref{fig:II_hl}, we show 
the Fourier transform of the oscillating part of the QFI density, 
$
        {f}_{Q,osc}^{(\mathrm{II})}(t) \equiv
        \frac{1}{V}
        \sum_\mathbf{k}
        \left[
        f_{ \mathbf{k}}^{(2)} \cos(2 E_{\kb}^{\infty}t)
        +
        f_{ \mathbf{k}}^{(4)} \cos(4 E_{\kb}^{\infty}t)
        \right]
$, 
for the long time asymptotic states following the representative phase II quenches ``a'', ``b'',  ``f'', ``g'', ``c'', ``d'', and ``h''.
These spectra can also be obtained from Fourier transforming the QFI density after subtracting its the long-time average,
${f}_{Q,osc}^{(\mathrm{II})}(t)=\frac{1}{V}({F}_Q^{(\mathrm{II})}(t)-\overline{F_Q^{(\mathrm{II})}})$,
or, equivalently, by removing the zero-frequency spike from the Fourier transform of the total QFI density $\frac{1}{V}{F}_Q^{(\mathrm{II})}(t)$.
The locations of these quenches in the phase diagram are shown in Fig.~\ref{fig:phase_diag},
and their quench coordinates ($\Delta_0^{(f)},\Delta_0^{(i)}$) are given in the corresponding figure captions.
The corresponding Cooper-pair distribution functions are shown in the insets.


As noted earlier, the minimum of the asymptotic quasiparticle energy $E_{\kb}^{\infty}$ occurs at $k=0$ when $\mu_{\infty} < m\Delta_{\infty}^2$ and moves to a non-zero momentum $k^*$ when $\mu_{\infty} > m\Delta_{\infty}^2$. These two regimes, separated by the green line $\mu_{\infty} = m\Delta_{\infty}^2$ in phase diagram Fig.~\ref{fig:phase_diag}, can be distinguished by the behaviour of the QFI spectrum at the continuum edge: in the former case, the QFI spectrum vanishes at the edge, whereas in the latter case, it instead exhibits a van Hove singularity.
As a concrete example, the phase II BCS-to-BCS quench ``a'' and BEC-to-BCS quench  ``b'' in Fig.~\ref{fig:II_ab} both satisfy $\mu_\infty>m\Delta_\infty^2 $
(i.e., lie on the left hand side of the $\mu_\infty=m\Delta_\infty^2$ line in Fig.~\ref{fig:phase_diag}), and their QFI spectra exhibit van Hove singularities at the continuum edges as predicted.
On the other hand, the BCS-to-BCS quench ``f'', BEC-to-BCS quench``g'' in Fig.~\ref{fig:II_fg}, together with the BEC-to-BEC quench ``c'', and BCS-to-BEC quench``d'' in Fig.~\ref{fig:II_cd} all satisfy $\mu_\infty<m\Delta_\infty^2$
(lie on the right hand side of $\mu_\infty=m\Delta_\infty^2$ line), and as expected their QFI spectra start continuously from zero at the continuum edges.

For phase II quenches with $\mu_\infty < m\Delta_\infty^2$, 
their QFI spectra, which start continuously from zero at edges, provide a direct probe of the topology of the pre-quench state. 
Specifically, although the QFI is computed for the post-quench asymptotic state, its behaviour near the continuum edge retains information about the pre-quench state carried by the pseudospin winding number $Q$.
We illustrate this using four representative phase II quenches, all satisfying $\mu_{\infty}<m\Delta_{\infty}^2$ (lying to the right of the $\mu_{\infty}=m\Delta_{\infty}^2$ line): the BCS-to-BCS quench ``f'' and  the BEC-to-BCS quench ``g'' shown in Fig.~\ref{fig:II_fg},  as well as the BEC-to-BEC quench ``c'' and the BCS-to-BEC quench ``d'' in Fig.~\ref{fig:II_cd}. 
We first compare quenches ``f'' and ``g'', both of which lie to the left of the $\mu_{\infty}=0$ line ($\mu_{\infty}>0$). Both quenches have the same post-quench topology characterized by the Green's function winding number  $W=1$, but different pre-quench topology characterized by the pseudospin winding number $Q$. 
Consistent with the analytical prediction, the QFI spectrum takes positive values immediately inside the continuum for quench ``f'', corresponding to the topological initial state with $Q=1$, whereas it take negative values for quench ``g'', corresponding to the trivial initial state with $Q=0$.
The same behaviour is observed for quenches ``c'' and ``d'' in Fig.~\ref{fig:II_cd}, both of which have the winding number $W=0$ but different $Q$. For the BEC-to-BEC quench ``c'' with $W=Q$, the distribution function $\gamma_{k}$ begins and ends at $\gamma_{k}=-1$ at both  $k=0$ and $k\to\infty$. By contrast, for the BCS-to-BEC quench ``d'' with $W\neq Q$, it changes from $+1$ at $k=0$ to $-1$ at $k\to\infty$.
The QFI spectrum again correctly identifies the topology of the pre-quench state.
Specifically, the sign of the QFI spectrum just inside the continuum near the edge $\mathrm{sgn}\tilde{F}_{Q,\mathrm{osc}}(\omega\rightarrow\pm (2E_0+0^+)) $ is negative for quench ``c'' with $Q=0$ and positive for quench ``d'' with $Q=1$. 

Note also that, for quenches ``f'' and ``c'', where $W=Q$, the distribution function approaches $\gamma_k=-1$ at both $k=0$ and $k\to\infty$. By contrast, for quenches ``g'' and ``d'', where $W\neq Q$, the distribution function winds from $\gamma_k=+1$ at $k=0$ to $\gamma_k=-1$ as $k\to\infty$.
While the Cooper-pair distribution function $\gamma_{\kb}$ can distinguish whether the same bulk topology of the pre-quench state is preserved after the quench (i.e., whether $Q=W$), it does not determine the explicit value of $Q$, which is instead encoded in the QFI spectrum.

Finally, we consider the QFI spectra for phase II quenches with very weak initial pairing strengths, $\Delta_i \ll \Delta_\mathrm{QCP}$. In this limit, the initial BCS state is similar to a normal Fermi sea with a sharp change in the pseudospin configuration at the Fermi momentum $k_F$. As a result, the
Cooper-pair distribution function $\gamma(k)$ becomes approximately a step function, taking the values
$\gamma(k) \approx \pm 1$ away from the step located at the Fermi momentum $k=k_F$~\cite{PRB,PRA}.
Since the continuum parts of the QFI spectrum receive contributions from $f^{(2)}_{\kb}$ and $f^{(4)}_{\kb}$, both of which vanish as $\gamma_\textbf{k}^2 \to 1$, only momenta in the vicinity of $k_F$  where $|\gamma(k\neq k_F)|$ deviates appreciably from $1$, contribute significantly to the continuum part of the QFI Fourier spectrum. 
Additionally, $f^{(2)}_{\kb} \propto \gamma_k\sqrt{1-\gamma_k^2}\Delta_{\infty}$ whereas $f^{(4)}_{\kb} \propto \left(1-\gamma_k^2\right)\Delta_{\infty}^2$. For phase II quenches with weak initial pairing, $\Delta_{\infty}$ is also small~\cite{PRB}, and as a result the $f^{(4)}_{\kb}$ contribution is strongly suppressed relative to the $f^{(2)}_{\kb}$ contribution. 
This results in pronounced peaks near frequencies $\omega=\pm 2E^{\infty} (k_F)$, while no comparable peaks appear near $\omega=\pm 4E^{\infty} (k_F)$, as demonstrated in the QFI spectrum for quench ``h'' in Figure \ref{fig:II_hl}.

\subsection{QFI in phase III}\label{sec:FIII}

In phase III, the asymptotic order parameter $\Delta_{\infty}(t)$ becomes periodic in time with period $T$, which is determined by the quench coordinates (explicit expression for $\Delta_{\infty}(t)$ is given in Refs.~\cite{PRB,PRA}). Consequently, the coherence factor
$\begin{Bmatrix}u_{\kb}(t) & v_{\kb}(t) \end{Bmatrix}^T$ is given by a superposition of two Floquet states related by particle-hole symmetry, with the corresponding weights determined by the Cooper-pair distribution function $\gamma_{\kb}$ [see Eq.~\ref{eq:uv}].

Since the coherence factors of these Floquet states, $\begin{Bmatrix}\tilde{u}_{\kb}(t) & \tilde{v}_{\kb}(t) \end{Bmatrix}^T$, are periodic in time (see Ref.~\cite{PRA} for the explicit expressions), they can be expanded in Fourier series as
\begin{equation}
	\begin{aligned}
		\tilde{u}_{\mathbf{k}}(t) = \sum_{n=-\infty}^\infty u_{\mathbf{k},n}e^{i\Omega n t},
		\qquad
		\tilde{v}_{\mathbf{k}}(t) = \sum_{n=-\infty}^\infty v_{\mathbf{k},n}e^{i\Omega n t}.
	\end{aligned}  
\end{equation}
Substituting these Fourier series into the QFI expression in Eq.~\ref{eq:QFI-uv} leads to the Fourier transform of the QFI in phase III
\begin{equation}\label{eq:FQ-III}
\begin{aligned}
    \tilde{f}_Q^{\mathrm{(III)}}(\omega) = &
    \frac{1}{V}
    \sum_{n=-\infty}^{\infty}
     \sum_{a=0,2,4}
    \sum_{\mathbf{k}}
    \pi\left[
    f^{(a)}_{n,\vex{k}}\delta(\omega-a E_\mathbf{k}^{(F)}-n\Omega)
    \right.
    \\
    &+\left.
    (f^{(a)}_{n,\vex{k}})^*
    \delta(\omega+a E_\mathbf{k}^{(F)}+n\Omega)
    \right],
 \end{aligned}   
\end{equation}
where
\begin{equation}\label{eq:f-iii}
\begin{aligned}
      &  f_{n,\mathbf{k}}^{(0)} = \sum_{n_1,n_2,n_3} 8\gamma_\mathbf{k}^2
\left[u^*_{\mathbf{k},n_1}u_{\mathbf{k},n_2}v^*_{\mathbf{k},n_3}v_{\mathbf{k},n +n_1-n_2+n_3}\right]\\
&+2(1-\gamma_\mathbf{k}^2)\sum_{n_1,n_2,n_3}
\left[u_{\mathbf{k},n_1}^*u_{\mathbf{k},n_2}u^*_{\mathbf{k},n_3}u_{\mathbf{k},n +n_1-n_2+n_3}\right.
\\
&\qquad\qquad\qquad\qquad
 \left.+v_{\mathbf{k},n_1}^*v_{\mathbf{k},n_2}v^*_{\mathbf{k},n_3}v_{\mathbf{k},n +n_1-n_2+n_3}\right],
\\
   &  f^{(2)}_{n,\vex{k}} =-\!\!\!\sum_{n_1, n_2,n_3}  \!\!\!8\gamma_\mathbf{k}\sqrt{1-\gamma_\mathbf{k}^2}
    \left[
u^*_{\mathbf{k},n_1}u_{\mathbf{k},n_2}u_{\mathbf{k},n_3}v_{\mathbf{k},n +n_1-n_2-n_3} 
    \right.
    \\
    &\qquad\qquad\qquad\qquad\left.
    -v^*_{\mathbf{k},n_1}v_{\mathbf{k},n_2}u_{\mathbf{k},n_3}v_{\mathbf{k},n +n_1-n_2-n_3}\right],
    \\
    &f^{(4)}_{n,\vex{k}} =  -\!\!\!\!\!\!
    \sum_{n_1, n_2,n_3} \!\!\! 4\left(1-\gamma_\mathbf{k}^2\right)u_{\mathbf{k},n_1}v_{\mathbf{k},n_2}u_{\mathbf{k},n_3}v_{\mathbf{k},n -n_1 -n_2 -n_3}.
    \end{aligned}
\end{equation}

Eq.~\ref{eq:FQ-III} shows that the QFI spectrum in phase III consists of two distinct contributions. The $a=0$ term gives rise to a series of evenly spaced peaks located at frequencies $\omega=n\Omega$, where $\Omega=2\pi/T$ is the order parameter oscillation frequency~\cite{PRA} and $n\in \mathbb{Z}$. Similar to the zero-frequency spike in phase II QFI spectrum, each of these peaks receives contributions from pseudospins at all momenta. Together they correspond to the periodic component of the QFI in the time domain. By contrast, the $a=2$ and $a=4$ terms, similar to their counterparts in phase II, contribute to continua in the QFI spectrum. Specifically, in the reduced zone representation, the Floquet quasienergy $E_{\kb}^{(F)}\in \left[-\Omega/2,+\Omega/2\right]$, and each pair of $\left(a,n\right)$ in Eq.~\ref{eq:FQ-III} contribute to the continua spanning the frequency intervals $\omega \in \left[(\pm n-a/2)\Omega,(\pm n+a/2)\Omega\right]$.  These continua generally overlap with each other. Moreover, the spectral weight of the continuum at a given frequency $\ww$ receives contributions from pseudospins whose Floquet quasienergies satisfy $\omega=\pm (aE_{\kb}^{(F)}+n\Omega)$. 
Therefore, when the frequency selects a local extremum of the Floquet quasienergy, the spectrum can exhibit a van Hove singularity. Unlike in phase II, these singularities are not generally located at the boundaries of the continua.

\begin{figure}[tbp]
    \centering
    \includegraphics[width=0.9\columnwidth]{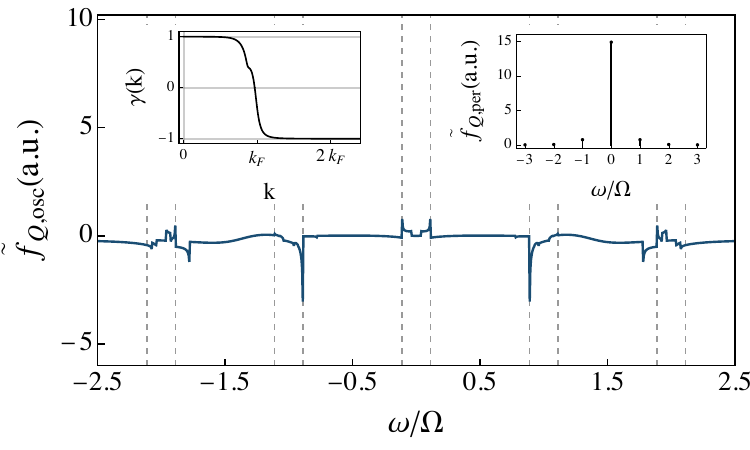}
    \caption{
    Fourier transform of the QFI density for the phase III quench ``A", with quench coordinate $(\Delta_0^{(i)}, \Delta_0^{(f)})=(0.12,0.51)\Delta_\mathrm{QCP}$. The main panel shows the continuum contribution arising from $a=2$ and $a=4$ terms in Eq.~\ref{eq:FQ-III}. The gray dashed lines indicate the frequencies $\ww=\pm 2E^{(F)}(q^*)+n\Omega$, $n\in \mathbb{Z}$, where $q^*$ denotes a momentum at which the Floquet quasienergy $E^{(F)}(k)$ reaches a local extremum.
    The left inset shows the Cooper-pair distribution function $\gamma(k)$. The right inset shows the $a=0$ contribution in Eq.~\ref{eq:FQ-III} at discrete frequencies $\ww=n\Omega$.  The resulting series of equally spaced discrete spikes at integer multiples of oscillation frequency is a characteristic of phase III Floquet dynamics.
    }
    \label{fig:III_A}
\end{figure}

In Fig.~\ref{fig:III_A}, we illustrate the QFI Fourier spectrum $\tilde{F}_Q(\omega)$ for a representative phase III quench ``A", which lies close to the phase II-III boundary (see Fig.~\ref{fig:phase_diag}). 
The main panel shows the combined continuum contribution from the $a=2$ and $a=4$ terms in Eq.~\ref{eq:FQ-III}, while the right inset shows the discrete peaks at integer multiples of the order parameter oscillation frequency from the $a=0$ terms. The right inset shows the weights of the discrete delta function peaks at $\ww=n\Omega$, whereas the main panel shows the continuum spectral weight. Their magnitudes therefore should not be compared directly.  The left inset shows the corresponding Cooper pair distribution function $\gamma(k)$, which enters the QFI spectrum through the weights $f^{(a)}_{n,\kb}$ in Eq.~\ref{eq:f-iii}.
The continuum spectrum shown in the main panel exhibits several van Hove singularities at frequencies $\ww=\pm 2E_{q^*}^{(F)}+n\Omega$, arising from the $a=2$ contribution. These frequencies are indicated by the gray dashed lines, and $q^*$ denotes the momentum at which the Floquet quasienergy reaches the local extremum. Two additional visible van Hove singularities, arising from the $a=4$ contribution, appear at frequencies $\ww=\pm 4 E_{q^*}^{(F)}$.
For the remaining $a=0$ contribution shown in the right inset, we note that the magnitude of the spectral weights of the discrete spikes located at frequencies $\ww=n\Omega$ decrease rapidly with increasing $|n|$. For quench "A" close to the phase II-III boundary, the Fourier components of the Floquet coherence factors, $u_{\kb,n}$ and $v_{\kb,n}$, decay rapidly with increasing $|n|$. This accounts for the rapid decrease in the spectral weight. 
For the same reason, when generating Fig.~\ref{fig:III_A}, we truncate the summation over Fourier components  in Eqs.~\ref{eq:FQ-III} and~\ref{eq:f-iii} at $|n|\leq 3$.

\subsection{Beyond the large-$L_A$ approximation}\label{sec:FC}

The calculations above make use of the condition that the subsystem linear size is much larger than the inverse of the characteristic momentum scale over which the pseudospin texture varies appreciatively, $L_A\gg k_s^{-1}$. Under this condition, the two pseudospins appearing in the exact expression Eq.~\ref{eq:FQA2} are approximately the same, and this reduces the double momentum summations there to the single summation in Eq.~\ref{eq:QFIA3}. We refer to this approximation below as the large-$L_A$ approximation. We now discuss the situation beyond the large-$L_A$ approximation, when this condition is relaxed (when $L_A\lesssim k_s^{-1}$), using the exact QFI expression in Eqs.~\ref{eq:FQA2} and~\ref{eq:FQA}.

In equilibrium, the ground-state QFI density associated with $N_A$ for any subsystem with $L_A \ll L$ can be obtained by inserting the equilibrium pseudospin configuration in Eq.~\ref{eq:s0} into the exact QFI expression in Eq.~\ref{eq:FQA2}:
\begin{align}
\begin{aligned}
    &f_Q^{0}
    =\frac{1}{V_A}
    \sum_{\kb,\kb'}
    |g(\kb-\kb')|^2 K(\kb,\kb'),
    \\
    &K(\kb,\kb')
    =
    1-
    \frac{\xi_{\kb}\xi_{\kb'}}{E_{\kb}E_{\kb'}}
    +
    \frac{\Delta_0^2 \kb\cdot\kb'}{E_{\kb}E_{\kb'}}.
\end{aligned}
\end{align}
Here $K(\kb,\kb')$ is bounded by $0\leq K(\kb,\kb')\leq 2$, and it varies continuously with $\mu_0$ at all momenta except for $\kb=0$ or $\kb'=0$. These points, however, carry vanishing measure in the double momentum integral. Therefore, we expect that the equilibrium QFI remains continuous across the topological quantum phase transition at $\mu_0=0$, even beyond the large-$L_A$ approximation.

In phase I, the pseudospin component $\braket{{s}_{\kb}^z}$ is static, while $\braket{{s}_{\kb}^-}$ oscillates at frequency $k^2/m$ (Eq.~\ref{eq:S-I}). As a result, for each pair of momenta $\left(\kb,\kb'\right)$, the term $\frac{1}{4} -\braket{{s}_{\kb}^z}\braket{{s}_{\kb'}^z}$ in Eq.~\ref{eq:FQA2} gives a time-independent contribution to the QFI $F_Q(t)$, corresponding to a zero-frequency peak in its Fourier spectrum $\tilde{F}_Q(\ww)$. In contrast, the term $\braket{{s}_{\kb}^+}\braket{{s}_{\kb'}^-}$ oscillates at frequency $(k^2-k'^2)/m$, contributing to  $\tilde{F}_Q(\ww)$ at frequency  
\begin{align}
\ww=\pm(k^2-k'^2)/m.
\end{align} 
Since the factor $|g(\kb-\kb')|^2$ restricts the double momentum summation to the regime $|\kb-\kb'| \lesssim L_A^{-1}$, this frequency reduces to approximately $\pm (k^2-k'^2)/m \approx \pm 2\kb \cdot (\kb-\kb')/m$ for large $L_A$. Additionally, the spectral weight of the oscillating term is proportional to  $\sqrt{(1-\gamma_{\kb}^2)(1-\gamma_{\kb'}^2)}$, which vanishes at sufficiently large momentum. 
The phase I QFI spectrum therefore consists of a static zero-frequency spike together with a narrow continuum centered at zero frequency, with width of the order of $O(v_c^{(\msf{I})}/L_A)$. Here we denote $k_c=mv_c^{(\msf{I})}$ the upper momentum cutoff below which $\gamma_k$ deviates appreciably from $-1$.
In the limit $L_A \rightarrow \infty$ and $L_A/L\to 0$, the continuum width vanishes, and the QFI spectrum reduces to a single zero-frequency spike,  recovering the time-independent phase-I result from large-$L_A$ approximation (see Sec.~\ref{sec:FI}).

In phase II, both $\braket{{s}_{\kb}^z}$ and $\braket{{s}_{\kb}^-}$ contain a static component as well as a component oscillating at frequency $2E_{\kb}^{\infty}$ (Eq.~\ref{eq:S-II}). The product of $\braket{\vex{s}_{\kb}}$ and $\braket{\vex{s}_{\kb'}}$ appearing in Eq.~\ref{eq:FQA2} therefore generates contributions to the QFI Fourier spectrum $\tilde{F}_Q(\omega)$ at frequency
\begin{align}
\ww=2(\sigma E_{\kb}^{\infty}+\sigma' E_{\kb'}^{\infty}),
\qquad
\sigma,\sigma'=0, \pm 1.
\end{align}
The term with $\sigma=\sigma'=0$ contributes to the zero frequency spike. Terms with $\sigma=\sigma'=\pm 1$ contribute for $|\omega|\geq 4E_0$.  Terms for which one of $\sigma$ and $\sigma'$ is zero and the other $\pm 1$ contribute for $|\omega|\geq 2E_0$.
Finally, terms with $\sigma=-\sigma'=\pm 1$ generate an additional continuum centered at zero frequency.
For $L_A$ large enough that energy difference can be approximated by $E_{\kb}^{\infty}-E_{\kb'}^{\infty} \approx \frac{\partial E_{\kb}^{\infty}}{\partial \kb}\cdot (\kb-\kb')$ over the momentum range
$|\kb-\kb'| \lesssim L_A^{-1}$,
the width of the central continuum is of the order $v_c^{(\msf{II})}/L_A$, where $v_c^{(\msf{II})}=\max_{k\leq k_c} |\frac{\partial E_{\kb}^{\infty}}{\partial \kb}|$.
Therefore, the QFI Fourier spectrum $\tilde{F}_Q(\ww)$ consists of the zero-frequency spike, the low-frequency continuum centered at it, together with positive- and negative-frequency continua occupying the regions $\ww\in [2E_{0},\infty)$ and $\ww\in (-\infty,-2E_{0}]$ respectively.
In the limit $L_A\rightarrow \infty$, as in phase I, the central continuum narrows and merges with the existing zero-frequency discrete spike, recovering the QFI spectrum obtained in Sec.~\ref{sec:FII} within the large-$L_A$ approximation.

We now consider the regime where the central continuum is separated by finite gaps from the positive and negative frequency continua, whose edge locations remain at $|\ww|=2E_0$. This is the case, for example, when $v_c^{(\msf{II})}/L_A \ll E_0$.
The edges for the left and right continua at frequency $|\ww|=2E_0$ receive only contributions from the products of the oscillating parts of the pseudospins at $k=k^*$, where $E_{k^*}^{\infty}=E_0$, and the static parts of the pseudospins at $\kb'$, with $|\kb-\kb'|\lesssim L_A^{-1}$.
When $\mu_{\infty}>m\Delta_{\infty}^2$, the minimum occurs at a finite momentum $k^*\neq 0$, and the QFI continues to exhibit van Hove singularities at the edges $|\ww|=2E_0$, as in the large-$L_A$ approximation.
In the opposite case $\mu_{\infty}<m\Delta_{\infty}^2$, the minimum occurs at $k^*=0$, so the pseudospin pairs contributing to the QFI at the edges $|\ww|=2E_0$ are those with $\kb=0$ and $\kb'$ running over $k'\lesssim L_A^{-1}$. Since the zero momentum pseudospin is frozen in time and possesses no oscillating part, the QFI $\tilde{F}_Q(\omega)$ at edges $|\ww|=2E_0$ vanish exactly, 
even for $L_A \lesssim k_s^{-1}$. 

Within the large-$L_A$ approximation, the sign of the QFI spectrum immediately inside the left and right continuum near the edges, i.e., $\sgn \tilde{F}_Q(\pm2(E_0+0^+))$,  encodes the pre-quench pseudospin winding number $Q$, as discussed in Sec.~\ref{sec:FII}. However, for finite $L_A$, $\sgn \tilde{F}_Q(\pm2(E_0+0^+))$ is no longer determined entirely by the zero momentum pseudospin alone, which carries the topological information of the pre-quench state. Instead it involves the pseudospin texture $\braket{\vex{s}_{\kb'}}$ over the momentum range $k'\lesssim L_A^{-1}$. $\sgn \tilde{F}_Q(\pm2(E_0+0^+))$ therefore depends on whether the signs of $\gamma_{\kb'}$ and $\xi_{\kb'}$ vary over this momentum window. A more complete analysis of $\sgn \tilde{F}_Q(\pm2(E_0+0^+))$, and whether it carries any topological information, outside the regime $L_A \gg k_s^{-1}$, requires evaluating the exact expression in Eq.~\ref{eq:FQA2} and is left for future study.

In phase III, the pseudospins have a  structure similar to the ones in phase II, but with the oscillation frequency replaced by the two times the Floquet quasienergy $2E_{\kb}^F$, and constant amplitude replaced by time periodic functions with the period identical to that of the order parameter (Eq.~\ref{eq:s-III}).
As a result, products of pairs of pseudospins at momentum $\kb$ and $\kb'$ appearing in Eq.~\ref{eq:FQA2} generate contributions to the QFI spectrum $\tilde{F}_Q(\ww)$ at frequencies
\begin{align}
\ww=2(\sigma E_{\kb}^{\msf{F}}+\sigma' E_{\kb'}^{\msf{F}})+n\Omega,
\quad
\sigma,\sigma'=0, \pm1,
\quad
n\in \mathbb{Z}.
\end{align}
In particular, the terms with $\sigma=\sigma'= 0$ generate discrete spikes at integer multiples of $\Omega$, as in the large-$L_A$ approximation.
Terms with $\sigma=-\sigma'\neq 0$ instead generate additional continua centered around these spike, with their widths of the order of $O(v_c^{(\msf{III})}/L_A)$ for large $L_A$. $v_c^{(\msf{III})}$ is defined analogously as its phase II counterpart $v_c^{(\msf{II})}$, i.e., $v_c^{(\msf{III})}=\max_{k\leq k_c} |\frac{\partial E_{\kb}^{F}}{\partial \kb}|$. 
The remaining choices of $\left(\sigma,\sigma'\right)$ produce continua whose boundaries are determined by the Floquet quasienergy spectrum and remain unchanged from those in the large-$L_A$ approximation. In the limit $L_A\rightarrow \infty$ with $L_A/L\rightarrow 0$, the continua centered on discrete spikes at $\ww=n\Omega$ narrows and merges with these discrete spikes, recovering the QFI spectrum obtained using the large-$L_A$ approximation in Sec.~\ref{sec:FIII}.


\section{Conclusion and outlook}\label{sec:conclusion}

In this work, we investigate the QFI associated with the particle number in a large and subextensive subsystem of a quenched 2D $p+ip$ superfluid. 
While the ground-state QFI does not provide a direct diagnostic of the equilibrium topological quantum phase transition, we show that its nonequilibrium counterpart contains richer information and, in particular, can reveal hidden topological information inaccessible in equilibrium. 
Within the large-$L_A$ approximation, $k_s^{-1}\ll L_A \ll L$, 
the QFI Fourier spectrum $\tilde{F}_Q(\ww)$ consists of a single zero-frequency spike in phase I. In phase II, in addition to the zero-frequency spike, the spectrum contains two additional continua occupying the region $|\ww|>2E_0$.  
In phase III, the spectrum contains a series of discrete spikes at integer multiples of the order parameter oscillation frequency $\Omega$, together with continua related to the Floquet quasienergy spectrum. 
These characteristic features persist  beyond the large-$L_A$ approximation when the condition $k_s^{-1}\ll L_A$ is relaxed. Each discrete spike - at zero frequency in phases I and II and at $\ww=n\Omega, n\in\mathbb{Z}$ in phase III - is now accompanied by an additional narrow continuum centered on the spike, with a width of the order of $O(v_c/L_A)$, which vanishes as $L_A\to \infty$. 
For the remaining continua in the QFI spectrum, their boundaries remain unchanged but their spectral weights are modified.
The arrangement of the discrete spikes and continua therefore continues to distinguish the three dynamical phases both within and beyond the large-$L_A$ approximation.
Additionally, in phase II, at the continuum edges $|\ww|=2E_0$, the QFI Fourier spectrum either exhibits van Hove singularities or vanishes exactly, depending on the sign of $(\mu_{\infty}-m\Delta_{\infty}^2)$,  and this behavior also persists when the large-$L_A$ condition is relaxed.
When the spectral weight  vanishes at the continuum edges ($m\Delta_{\infty}^2>\mu_{\infty}$), the sign of the QFI spectrum just inside the continuum close to the edge  $\sgn \tilde{F}_Q(\pm2(E_0+0^+))$ is positive for quenches from the topological BCS phase and negative for quenches from the trivial BEC phase for sufficiently large $L_A$.
These results demonstrate that driving the system out of equilibrium provides a route to enhancing the diagnostic power of the QFI associated with a simple physical observable, without the need to choose generators tailored to a particular order parameter or symmetry of the system. 


In the present work, we use the particle number in a large and subextensive subsystem as the generator, but we expect the nonequilibrium QFI to behave similarly for other physical observables. The nonequilibrium QFI is determined by both the post-quench quasiparticle spectrum and the Cooper pair distribution function, both of which carry topological information. It is this additional information that enables the QFI to access topology that is hidden in equilibrium.
Time dependence of the QFI is another important ingredient for this enhancement. The quench makes the QFI time dependent (except for phase I within the large-$L_A$ approximation), and as a result its Fourier transform resolves contributions from pseudospins at different momenta. In particular, it isolates the zero-momentum pseudospin, in which the pre-quench winding number is encoded. 
These mechanisms do not rely on the particular form of the generators, and we therefore expect an enhancement of the diagnostic power for QFI generated by other physical observables.
Here we focus on an integrable model, whose long-time asymptotic dynamics is known exactly in the thermodynamic limit, allowing the nonequilibrium QFI to be obtained analytically. It would be interesting to explore whether a similar enhancement of the diagnostic power of the QFI persists in generic many-body interacting systems, particularly those with chaotic dynamics~\cite{pappalardi2018,Pappalardi2020,Srednicki,bergman2026}. 

\section*{Acknowledgments}

YL acknowledges support from the Knut and Alice Wallenberg Foundation (KAW 2024.0131).

\appendix
\vspace{1em}

\section{QFI in equilibrium}~\label{app:eqQFI}
In this Appendix, we provide the full expression for the QFI density of the equilibrium ground state. Performing the integral in Eq.~\ref{eq:qfi0}, one finds a closed form expression whose leading term is given in Eq.~\ref{eq:qfi0-1}. The remaining $\mathcal{O}(1)$ contributions depend on the sign of $m\Delta_0^2 -2\mu_0$:
\begin{widetext}
\begin{equation}
    f_Q^{0} = \frac{2m^2\Delta_0^2}{\pi}\begin{cases}
        \frac{x_1+\mu_0}{x_1-x_2}\ln{\left|\frac{\Lambda-x_1}{-\mu_0-x_1}\right|}-\frac{x_2+\mu_0}{x_1-x_2}\ln{\left|\frac{\Lambda-x_2}{-\mu_0-x_2}\right|},& m\Delta_0^2 > 2\mu_0
    \\
    \frac{1}{2}\ln{\left|\frac{(\Lambda-\alpha)^2+\beta^2}{\mu_0^2}\right|}+
    \frac{\alpha+\mu_0}{\beta}\left\{\arctan\left[\frac{\Lambda-\alpha}{\beta}\right]- \arctan\left[\frac{-\mu_0-\alpha}{\beta}\right]\right\},&m\Delta_0^2 < 2\mu_0
    \\
    \ln{\left|\frac{\Lambda+m\Delta_0^2}{m\Delta_0^2-\mu_0}\right|}+\frac{\mu_0}{\Lambda+m\Delta_0^2}-1, & m\Delta_0^2 = 2\mu_0,
    \end{cases}
\end{equation}
Here $x_{1,2} = m\Delta_0^2\left[-1\pm\sqrt{1-\frac{2\mu_0}{m\Delta_0^2}}\right]$, $\alpha = -m\Delta_0^2$ and $\beta = \sqrt{2m\Delta_0^2\mu_0-m^2\Delta_0^4}$.
\end{widetext}

\bibliographystyle{apsrev4-2} 
\bibliography{references}      
        
	\end{document}